\pdfoutput=1
\PassOptionsToPackage{table,dvipsnames}{xcolor}
\PassOptionsToPackage{dvipsnames}{colortbl}
\documentclass{article}
\usepackage{iclr2027_conference,times}
\usepackage{etoolbox}
\usepackage{enumitem}
\usepackage{algorithm}
\usepackage{algpseudocode}
\floatstyle{plain}
\restylefloat{algorithm}
\usepackage{wrapfig}
\usepackage{needspace}
\algnewcommand{\LineComment}[1]{\State\textcolor{blue}{\(\triangleright\)#1}}

\usepackage{amsmath,amsfonts,bm}
\usepackage{hyperref}
\usepackage{url}
\usepackage{graphicx}
\usepackage{subcaption}
\usepackage{booktabs}
\usepackage{multirow}
\usepackage{tabularx}
\usepackage{arydshln}
\usepackage[table]{xcolor}
\usepackage{makecell}
\usepackage{listings}
\usepackage{tcolorbox}
\tcbuselibrary{listings,breakable}
\usepackage{capt-of}
\usepackage{tikz}
\usetikzlibrary{arrows.meta,positioning,calc,fit,backgrounds}

\definecolor{darkcerulean}{rgb}{0.03, 0.27, 0.49}
\definecolor{burgundy}{rgb}{0.7, 0.0, 0.13}
\definecolor{majorelleblue}{rgb}{0.38, 0.31, 0.86}
\definecolor{forestgreen}{rgb}{0.13, 0.55, 0.13}
\definecolor{arylideyellow}{rgb}{0.89, 0.61, 0.06}
\definecolor{arsenic}{rgb}{0.23, 0.27, 0.29}
\definecolor{GACloneBar}{HTML}{A3B7CA}
\definecolor{GAEncBar}{HTML}{476F95}

\lstdefinestyle{promptstyle}{
  basicstyle=\ttfamily\scriptsize\color{arsenic},
  columns=fullflexible,
  keepspaces=true,
  breaklines=true,
  showstringspaces=false,
  tabsize=2,
}

\newtcblisting{promptbox}{
  listing only,
  breakable,
  colback=white,
  colframe=black!15,
  boxrule=0.5pt,
  arc=2pt,
  left=6pt,right=6pt,top=6pt,
  listing options={style=promptstyle}
}

\hypersetup{
  colorlinks=true,
  citecolor=darkcerulean,
  linkcolor=burgundy,
  urlcolor=majorelleblue,
  pdftitle={Launder-Bench: Can Repository-Level Agentic Code Laundering Be Detected?},
  pdfauthor={Yuntong Hu, Hongji Pu, Zimo Li, Ruomeng Li, Xiaotong Liu, Ziyuan Qin, Qiyue Li, Liang Zhao}
}

\newcommand{\LinkTable}[1]{\hyperref[#1]{Table~\ref*{#1}}}
\newcommand{\LinkFigure}[1]{\hyperref[#1]{Figure~\ref*{#1}}}
\newcommand{\LinkSection}[1]{\hyperref[#1]{Section~\ref*{#1}}}
\newcommand{\LinkAppendix}[1]{\hyperref[#1]{Appendix~\ref*{#1}}}
\newcommand{\LinkAlgorithm}[1]{\hyperref[#1]{Algorithm~\ref*{#1}}}

\definecolor{TableHeader}{HTML}{F2F2F2}
\definecolor{TableFocus}{HTML}{EDE7F6}
\definecolor{TableInk}{HTML}{453A5C}
\newcolumntype{L}{>{\raggedright\arraybackslash}X}

\newcommand{\tablestyle}{%
  \footnotesize
  \renewcommand{\arraystretch}{1.18}%
  \setlength{\tabcolsep}{4.5pt}%
}
\newcommand{\tablehead}[1]{\textbf{#1}}
\newcommand{\tablegroup}[2]{%
  \hline\multicolumn{#1}{c}{\color{TableInk}\bfseries #2}\\}

\newcommand{\nSources}{282}
\newcommand{\nNegatives}{1{,}244}
\newcommand{\nNegMedianSim}{0.076}
\newcommand{\nVerified}{1{,}232}
\newcommand{\nOraclePassed}{1{,}243}
\newcommand{\nPair}{1{,}128}

\newcommand{\nAttr}{229}
\newcommand{\nPairPerSource}{4}
\newcommand{\nPairRelease}{1{,}128}
\newcommand{\nAttrEligible}{229}
\newcommand{\pairOkGoldenMin}{5}
\newcommand{\refOrigTokens}{11{,}000}
\newcommand{\nSourcesPy}{199}
\newcommand{\nSourcesJava}{83}

\newcommand{\benchmark}{Launder-Bench}
\newcommand{\benchpair}{\benchmark{}-Pair}
\newcommand{\benchattr}{\benchmark{}-Attr}

\title{\benchmark{}: Can Repository-Level\\Agentic Code Laundering Be Detected?}

\author{%
\parbox[t]{\dimexpr\textwidth-2\tabcolsep\relax}{%
  \centering
  \mbox{\textbf{Yuntong Hu$^{1}$}} \quad
  \mbox{\textbf{Hongji Pu$^{2}$}} \quad
  \mbox{\textbf{Zimo Li$^{1}$}} \quad
  \mbox{\textbf{Ruomeng Li$^{1}$}} \quad
  \mbox{\textbf{Xiaotong Liu$^{1}$}} \quad
  \mbox{\textbf{Ziyuan Qin$^{1}$}} \quad
  \mbox{\textbf{Qiyue Li$^{3}$}} \quad
  \mbox{\textbf{Liang Zhao$^{1,\dagger}$}}\endgraf
  \vspace{3pt}
  {\normalfont
    $^{1}$Emory University \quad
    $^{2}$University of Illinois Urbana-Champaign\endgraf
    $^{3}$Beijing Normal-Hong Kong Baptist University\endgraf
    \vspace{3pt}
    \mbox{\texttt{hongjip2@illinois.edu}} \quad
    \mbox{\texttt{t330005026@mail.bnbu.edu.cn}}\endgraf
    \mbox{\texttt{\{yuntong.hu,zimo.li,lina.li\}@emory.edu}}\endgraf
    \mbox{\texttt{\{steven.liu,ziyuan.qin,liang.zhao\}@emory.edu}}\endgraf
    \vspace{-20pt}
  }%
}%
}

\iclrfinalcopy

\makeatletter
\patchcmd{\@maketitle}
  {\lhead{Published as a conference paper at ICLR 2027}}
  {\lhead{Preprint}}
  {}
  {\PackageError{Launder-Bench}{Could not patch ICLR preprint header}{Check the iclr2027_conference.sty version}}
\makeatother

\begin{document}

\maketitle

\begin{abstract}
Coding agents can now rewrite repository-scale codebases while preserving tested
behavior, creating a new challenge for plagiarism detection, license auditing,
and software provenance. We ask whether such rewrites can still be traced to
their source. Existing clone benchmarks emphasize fragments and bounded edits,
while repository-transformation benchmarks evaluate functional correctness
rather than provenance. Moreover, ranking the most similar source does not
establish derivation when independent implementations can be similarly close.
We define \emph{repository-level agentic code laundering} and introduce
\benchmark{}, to our knowledge the first repository-level benchmark for
provenance under agentic rewriting, constructed from $\nSources$ real-world
repositories. A source-hidden reconstruction protocol decomposes source
behavior, reconstructs modules and repository structure, and retains rewrites
only after execution-based verification. We evaluate the same provenance
problem in two complementary settings: \benchpair{} contains
$\nPairRelease$ pairwise instances asking whether two repositories share
provenance, while \benchattr{} contains $\nAttrEligible$ source-attribution
instances asking which candidate is the source, with abstention when evidence
is insufficient. Measured same-domain independent repositories establish how
much similarity can arise without shared provenance and provide
detector-specific reference bands.
Existing methods remain far from saturating \benchmark{}: the strongest static
method reaches only $0.66$ AUROC on pairwise detection, while repository
encoders rank the true source first on up to $22\%$ of attribution queries but
earn calibrated attribution credit on only $1$--$2\%$. These results reveal a
systematic gap between relative code similarity and source-specific provenance
evidence, providing a reproducible target for provenance detection under
agentic rewriting. https://github.com/HuieL/Launder-Bench 
\end{abstract}

\begin{figure*}[h]
\centering
\begin{minipage}[t]{0.485\textwidth}
  \centering
  \includegraphics[width=\textwidth]{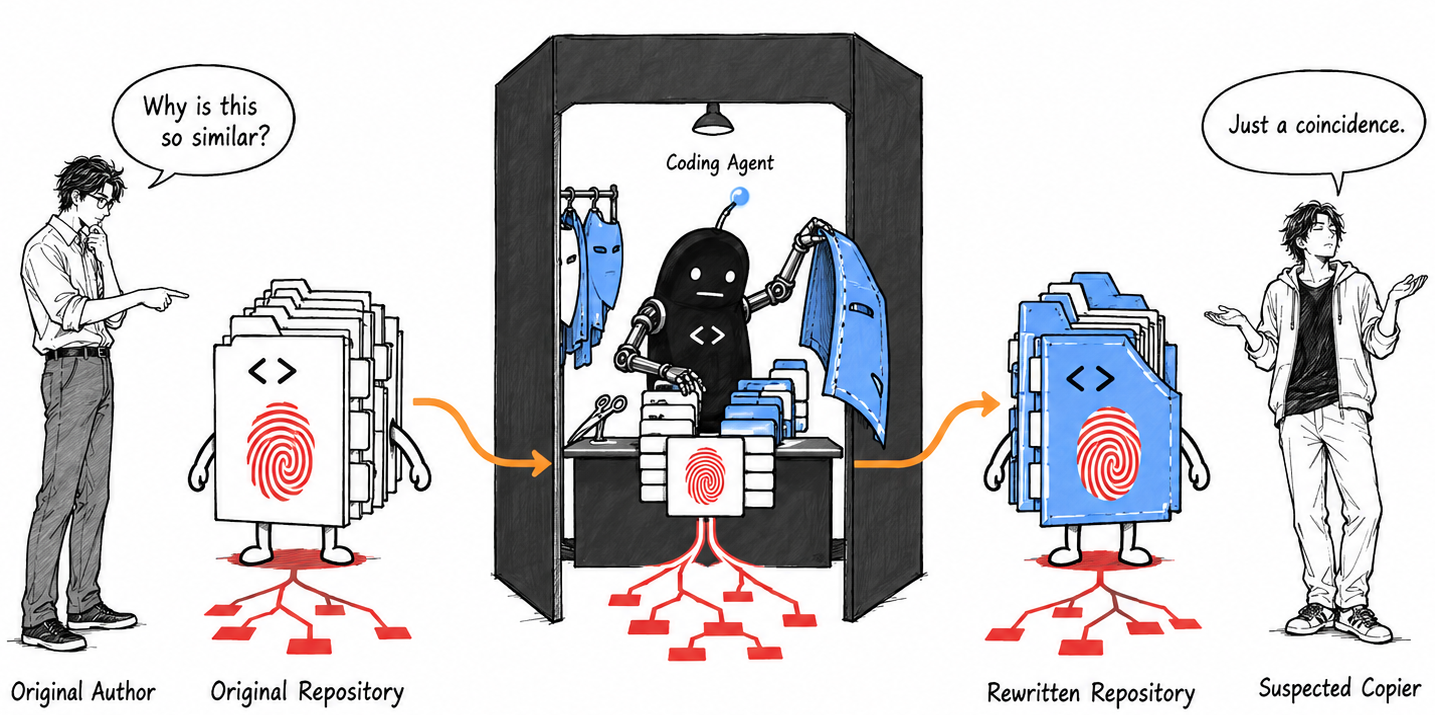}\\[2pt]
  \textbf{(a)} Repository-scale agentic rewriting
\end{minipage}\hfill
\begin{minipage}[t]{0.485\textwidth}
  \centering
  \includegraphics[width=\textwidth]{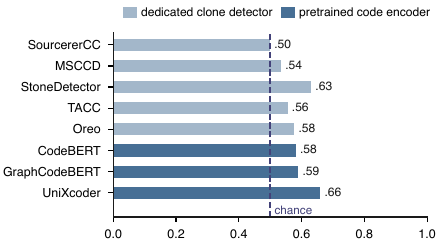}\\[2pt]
  \textbf{(b)} Pairwise provenance detection (AUROC)
\end{minipage}
\caption{
\textbf{(a)} Repository-level agentic code laundering: a coding agent transforms a
source repository into a rewrite that preserves tested behavior while
obscuring source-specific provenance evidence.
\textbf{(b)} Existing methods on our pairwise provenance detection task (AUROC).
{\color{GACloneBar}\textbf{Light bars}} denote dedicated clone detectors with supported-input
coverage; these records are provisional and have incomplete coverage. StoneDetector, TACC, and Oreo support only Java$\rightarrow$Java.
{\color{GAEncBar}\textbf{Dark bars}} denote pretrained code encoders evaluated on the full benchmark.
The dashed line indicates chance-level discrimination (AUROC $=0.5$).
The provenance relation remains even when detection is difficult.
}
\label{fig:graphical-abstract}
\end{figure*}

\section{Introduction}
\label{sec:intro}

Coding agents can now rewrite repository-scale code while preserving tested
behavior. Agentic rewriting can rename identifiers, restructure module
boundaries, reimplement algorithms, and translate implementations across
programming languages. Yet the resulting repository remains \emph{derived}:
these transformations can obscure source-specific provenance evidence without
removing the underlying provenance relation.

This creates a new threat we call \textbf{repository-level agentic code laundering}: producing a derived implementation that retains the tested capabilities of a codebase while reducing the similarity evidence used to distinguish derivation from independent work.
\LinkFigure{fig:graphical-abstract}\hyperref[fig:graphical-abstract]{a} illustrates the setting. It matters for plagiarism detection~\citep{aiken1994moss,prechelt2002finding}, license-compliance auditing~\citep{inoue2021code}, and software forensics, where similarity provides evidence of reuse.

Existing clone benchmarks primarily test fragments and bounded edits~\citep{svajlenko2014towards,roy2018benchmarks}, and adversarial perturbations test model robustness within existing programs~\citep{yefet2020adversarial}. 
Repository transformation benchmarks evaluate whether an agent can produce a functionally valid implementation \citep{wang2026repotransbench,li2026repomod}, but do not establish whether the resulting repository can still be traced to its source. 
Existing semantic-clone detectors degrade substantially under semantics-preserving transformations~\citep{xu2026semantic}, suggesting an even greater challenge under repository-scale agentic reconstruction.

Nearest-source ranking alone does not answer that question either.
Independently developed repositories in the same domain can share interfaces,
vocabulary, and implementation patterns. A detector may rank the true source
first without distinguishing it from such independent code. Provenance
evaluation therefore needs behavior-verified rewrites, controlled non-source
comparisons, and an evidence requirement that permits abstention.

We introduce \benchmark{}, a repository-level provenance benchmark with
execution-verified rewrites across same-language and cross-language axes.
\benchpair{} tests shared provenance between repositories; \benchattr{}
tests source recovery among candidates with an evidence requirement.
These are complementary views of the same provenance records. Measured
same-domain independent repositories provide a source- and detector-specific
reference for similarity without known derivation. This design separates relative retrieval from evidence that exceeds the
independent-code reference.

Our contributions are as follows:
\begin{itemize}[left=0pt]
\item \emph{Problem formulation} (\LinkSection{sec:tasks}). We formulate repository-level agentic code laundering as a provenance problem: derivation persists even when rewriting obscures the evidence available to a detector.
\item \emph{Benchmark} (\LinkSection{sec:task-construction}). We construct complementary Pair and Attribution tasks from behavior-verified agentic rewrites, with measured same-domain references and fixed candidate sets.
\item \emph{Empirical findings} (\LinkSection{sec:detectors}). The strongest full-coverage static detector reaches AUROC $0.661$ and best F1 $0.394$. Encoders retain up to $22\%$ Rank@1 but only $1$--$2\%$ calibrated attribution, exposing a gap between relative similarity and source-specific evidence.
\end{itemize}

\section{\benchmark{}}
\label{sec:benchmark}

\benchmark{} evaluates provenance after coding agents transform real-world
repositories while preserving tested behavior. Pairwise detection and source
attribution provide complementary views of this problem.

\subsection{Task Formulation}
\label{sec:tasks}

Let $\pi$ denote an agentic rewriting policy and $\mathcal{P}_q$ the provenance set of source repositories provided to it. The policy produces a rewritten repository $R_q \sim \pi(\cdot \mid \mathcal{P}_q)$ whose tested functionality is preserved. The released benchmark instantiates the single-source setting, where $|\mathcal{P}_q|=1$.

\paragraph{Pairwise provenance detection.}
Let $\mathcal{P}(R)$ denote a repository's source-ancestor set in the
benchmark provenance records. For an original source $R_s$,
$\mathcal{P}(R_s)=\{R_s\}$; for a rewrite $R_q$,
$\mathcal{P}(R_q)=\mathcal{P}_q$. Given $(R_q,R_c)$, the task is to determine
whether $\mathcal{P}(R_q)\cap\mathcal{P}(R_c)\neq\emptyset$.
This includes direct source--rewrite pairs and rewrite--rewrite pairs with a
shared source. A detector assigns a provenance score $d(R_q,R_c)$, where a
larger score indicates stronger evidence of shared provenance.
We evaluate pairwise discrimination using AUROC and best F1, with decision
metrics detailed in \LinkAppendix{app:metrics}.

\paragraph{Source attribution.}
Given $R_q$ and a candidate corpus $\mathcal{C}$, the task is to recover the source in $\mathcal{P}_q\subseteq\mathcal{C}$ by ranking candidates according to $d(R_q,R_c)$. We evaluate source retrieval and whether its evidence exceeds the source-specific independent-code reference band, as detailed in
\LinkAppendix{app:metrics}.

\subsection{Task Construction}
\label{sec:task-construction}

We construct both tasks from verified rewrites and label them by provenance
paths rather than surface similarity. The design varies provenance, functionality, and domain relatedness so that a detector cannot succeed merely by recognizing the task or programming language.

\Needspace{20\baselineskip}
\begin{figure}[t]
\centering
\begingroup%
\colorlet{rewriteSource}{darkcerulean}%
\colorlet{rewriteTarget}{TableInk}%
\colorlet{rewriteNeutral}{black!65}%
\begin{tikzpicture}[
  x=\linewidth,y=1mm,
  font=\fontsize{8}{9}\selectfont,text=black!90,
  line cap=round,line join=round,
  heading/.style={font=\fontsize{8}{9}\selectfont\bfseries,
    align=center,inner sep=0pt},
  stageheading/.style={heading,font=\fontsize{7.3}{8.2}\selectfont\bfseries},
  artifact/.style={draw,line width=.55pt,rounded corners=1.5pt,
    fill=white,minimum height=22mm,inner sep=0pt},
  repo/.style={artifact,minimum width=.12\linewidth},
  process/.style={draw=rewriteNeutral,line width=.55pt,rounded corners=1.5pt,
    fill=white,align=center,inner sep=1.5pt},
  spec/.style={draw=rewriteNeutral,line width=.45pt,fill=white,
    rounded corners=.7pt,minimum width=5mm,minimum height=4.5mm,
    inner sep=.4pt,font=\fontsize{9}{10}\selectfont},
  module/.style={spec,draw=rewriteTarget,fill=rewriteTarget!7,
    minimum width=6.5mm},
  flow/.style={-{Stealth[length=1.3mm,width=.9mm]},
    draw=black!65,line width=.6pt},
  dependency/.style={-{Stealth[length=.7mm,width=.6mm]},
    draw=black!50,line width=.4pt},
  feedback/.style={flow,draw=black!55,line width=.45pt,dashed,
    rounded corners=1mm},
  file/.style={font=\fontsize{7}{8}\selectfont\ttfamily,
    anchor=base west,inner sep=0pt,text height=1.9mm,text depth=.5mm},
  note/.style={font=\fontsize{6.5}{7.2}\selectfont,align=center,inner sep=.5pt}
]
\coordinate (mainY) at (0,0);
\coordinate (sourcecol) at (.060,0);
\coordinate (speccol) at (.198,0);
\coordinate (targetspecscol) at (.351,0);
\coordinate (agentcol) at (.520,0);
\coordinate (repocol) at (.688,0);
\coordinate (verifycol) at (.836,0);
\coordinate (verifiedcol) at (.958,0);
\node[repo,draw=rewriteSource] (source) at (sourcecol |- mainY) {};
\node[artifact,draw=black!30,minimum width=.125\linewidth]
  (specs) at (speccol |- mainY) {};
\node[artifact,draw=rewriteTarget!65,minimum width=.125\linewidth]
  (targetspecs) at (targetspecscol |- mainY) {};
\node[artifact,draw=rewriteTarget!65,minimum width=.175\linewidth]
  (agent) at (agentcol |- mainY) {};
\node[repo,draw=rewriteTarget] (rewrite) at (repocol |- mainY) {};
\node[process,minimum width=.12\linewidth,minimum height=10mm]
  (verify) at (verifycol |- mainY) {Behavioral\\Verification};
\node[process,draw=rewriteTarget,text=rewriteTarget,fill=rewriteTarget!8,
  font=\fontsize{7.5}{8.5}\selectfont\bfseries,
  minimum width=.083\linewidth,minimum height=9mm]
  (verified) at (verifiedcol |- mainY) {Verified\\Rewrite};

\node[stageheading,text=rewriteSource!80] (h1) at (.135,21.5)
  {1\quad SOURCE\\UNDERSTANDING};
\node[stageheading,text=rewriteTarget!80] (h2) at (.444,21.5)
  {2\quad SOURCE-HIDDEN\\RECONSTRUCTION};
\node[stageheading,text=black!65] (h3) at (.812,21.5)
  {3\quad EXECUTION-BASED\\VERIFICATION};

\foreach \name/\title/\col in {source/Source/rewriteSource,rewrite/Rewritten/rewriteTarget}{
  \node[heading,text=\col] at ([yshift=7.4mm]\name.center)
    {\title\\Repository};
}
\node[file] (srcroot) at ([xshift=1mm,yshift=.5mm]source.west) {src/};
\node[file] (srca) at ([xshift=1.5mm,yshift=-3mm]srcroot.base west) {parser.py};
\node[file] (srcb) at ([yshift=-3mm]srca.base west) {crypto.py};
\node[file] (srcc) at ([yshift=-3mm]srcb.base west) {utils.py};
\node[file] (newroot) at ([xshift=1mm,yshift=.5mm]rewrite.west) {core/};
\node[file] (newa) at ([xshift=1.5mm,yshift=-3mm]newroot.base west) {engine.py};
\node[file] (newb) at ([yshift=-3mm]newa.base west) {codec.py};
\node[file] (newc) at ([yshift=-3mm]newb.base west) {aux.py};
\foreach \root/\first/\second/\last/\col in {
  srcroot/srca/srcb/srcc/rewriteSource,newroot/newa/newb/newc/rewriteTarget}{
  \draw[\col!65,line width=.35pt]
    ([xshift=.4mm,yshift=-.6mm]\root.south west)
    |- ([xshift=-.6mm]\last.west);
  \draw[\col!65,line width=.35pt]
    ([xshift=-1.1mm]\first.west) -- ([xshift=-.6mm]\first.west)
    ([xshift=-1.1mm]\second.west) -- ([xshift=-.6mm]\second.west);
}
\draw[rewriteSource,line width=.5pt,fill=rewriteSource!9]
  ([xshift=1mm]source.north west) -- ++(0,1mm) -- ++(6mm,0)
  -- ++(1mm,-1mm) -- cycle;

\node[heading] at ([yshift=7.4mm]specs.center) {Module\\Specifications};
\node[spec] (s1) at ([yshift=1.5mm]specs.center) {$S_1$};
\node[spec] (s2) at ([xshift=-4.3mm,yshift=-4mm]specs.center) {$S_2$};
\node[spec] (s3) at ([xshift=4.3mm,yshift=-4mm]specs.center) {$S_3$};
\draw[dependency] (s1) -- (s2);
\draw[dependency] (s1) -- (s3);
\node[note] at ([yshift=-8.5mm]specs.center) {DAG \& interfaces};

\node[heading,text=rewriteTarget] at ([yshift=7.4mm]targetspecs.center)
  {Rewritten\\Specifications};
\node[spec,draw=rewriteTarget] (t1) at ([xshift=-4.1mm,yshift=1.5mm]targetspecs.center) {$T_1$};
\node[spec,draw=rewriteTarget] (t2) at ([xshift=4.1mm,yshift=1.5mm]targetspecs.center) {$T_2$};
\node[spec,draw=rewriteTarget] (t3) at ([yshift=-3.5mm]targetspecs.center) {$T_3$};
\draw[dependency] (t1) -- (t3);
\draw[dependency] (t2) -- (t3);
\node[note] at ([yshift=-8.5mm]targetspecs.center) {New DAG\\\& interfaces};

\node[heading,text=rewriteTarget] at ([yshift=8.7mm]agent.center) {Coding Agent};
\node[note,font=\fontsize{6.3}{7}\selectfont] at ([yshift=5.9mm]agent.center)
  {Module reconstruction};
\foreach \i/\rowY in {1/2,2/-3.3,3/-8.6}{
  \node[spec] (a\i) at ([xshift=-7.2mm,yshift=\rowY mm]agent.center) {$T_\i$};
  \node[module] (m\i) at ([xshift=2.2mm,yshift=\rowY mm]agent.center) {$M'_\i$};
  \draw[flow] (a\i) -- (m\i);
}
\draw[dependency] (m2) -- (m3);
\draw[dependency,rounded corners=.6mm] (m1.east) -- ++(1.5mm,0) |- (m3.east);
\node[note,rotate=90,anchor=north] at ([xshift=2.1mm]m2.east) {interfaces};

\draw[flow] (source.east) -- (specs.west);
\draw[flow] (specs.east) -- (targetspecs.west);
\draw[flow] (targetspecs.east) -- (agent.west);
\draw[flow] (agent.east) -- (rewrite.west);
\draw[flow] (rewrite.east) -- (verify.west);
\draw[flow,draw=rewriteTarget] (verify.east)
  -- node[above,note,yshift=5mm] {pass} (verified.west);

\coordinate (decompose) at ($(source.east)!.5!(specs.west)$);
\coordinate (assembly) at ($(agent.east)!.5!(rewrite.west)$);
\node[note,text=rewriteSource,anchor=north] (decomposelabel)
  at (decompose |- {$(mainY)+(0,-12mm)$}) {Behavioral\\Decomposition};
\draw[black!35,line width=.35pt] (decomposelabel.north) -- ([yshift=-.9mm]decompose);
\node[note,text=rewriteTarget,anchor=north] (assemblylabel)
  at (assembly |- {$(mainY)+(0,-12mm)$}) {Repository\\Assembly};
\draw[black!35,line width=.35pt] (assemblylabel.north) -- ([yshift=-.9mm]assembly);

\node[process,draw=rewriteSource!65,text=rewriteSource,
  font=\fontsize{7}{8}\selectfont] (golden) at (.16,15.5)
  {Golden Behavioral Checks};
\draw[flow,draw=rewriteSource!70,rounded corners=.8mm]
  (decompose) |- ([yshift=-1mm]golden.south) -- (golden.south);
\draw[flow,draw=rewriteSource!70,rounded corners=1mm]
  (golden.east) -| (verify.north);

\coordinate (boundary) at (.275,0);
\draw[black!60,densely dashed,line width=.6pt]
  ([yshift=10mm]boundary) -- ([yshift=-10mm]boundary);
\node[note,font=\fontsize{6.5}{7.2}\selectfont\bfseries,anchor=north]
  at ([yshift=-12mm]boundary) {Source Implementation\\Hidden};

\draw[feedback] (verify.south) -- (verifycol |- {$(mainY)+(0,-17mm)$})
  -| (agent.south);
\node[note,anchor=south] at (.735,-16.2) {fail: Execution Feedback};
\coordinate (stageTop) at ([yshift=.4mm]h1.north);
\coordinate (stageBottom) at ([yshift=-.5mm]source.south);
\begin{scope}[on background layer]
  \path[fill=rewriteSource!3,rounded corners=1.5pt]
    (0,0 |- stageTop) rectangle (.265,0 |- stageBottom);
  \path[fill=rewriteTarget!4,rounded corners=1.5pt]
    (.285,0 |- stageTop) rectangle (.618,0 |- stageBottom);
  \path[fill=black!2,rounded corners=1.5pt]
    (.625,0 |- stageTop) rectangle (1,0 |- stageBottom);
\end{scope}
\end{tikzpicture}%
\endgroup%
\caption{Agentic repository transformation. The source is decomposed into
behavioral module specifications and executable checks, then hidden; a coding
agent reconstructs the repository from restructured specifications and repairs
it under execution feedback. Only rewrites that pass the behavioral checks are
retained. DAGs and file names are schematic.}
\label{fig:agentic-rewriting}
\end{figure}
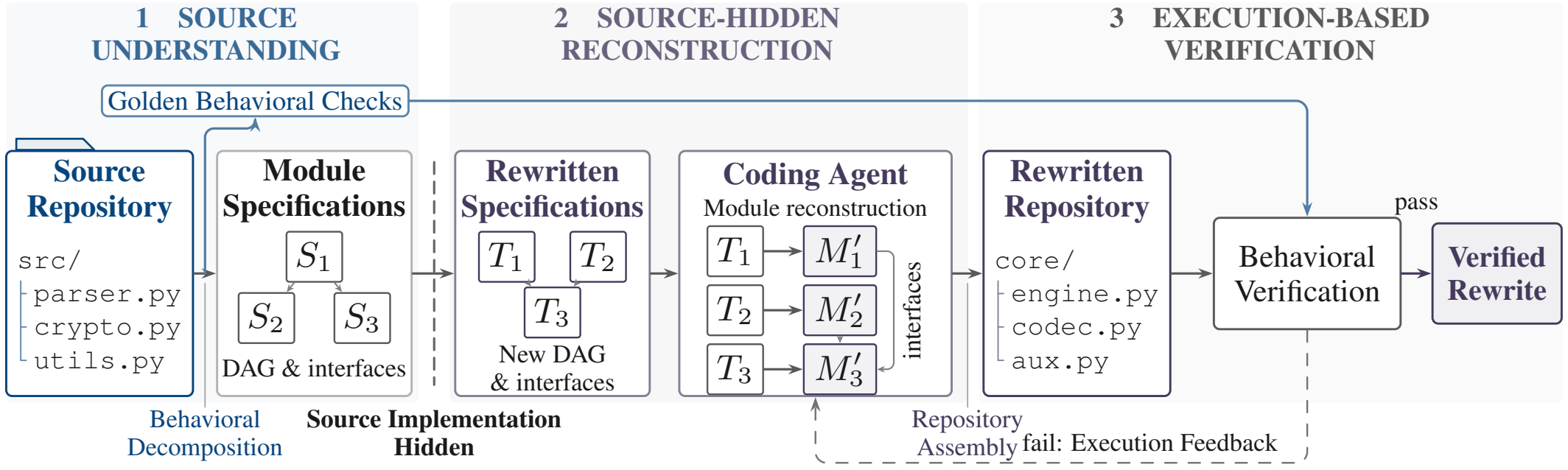
\Needspace{24\baselineskip}
\begin{wrapfloat}{algorithm}[23]{r}[0pt]{0.51\columnwidth}
\vspace{-\baselineskip}
\hrule height 0.8pt
\vspace{2pt}
\refstepcounter{algorithm}
\textbf{Algorithm \thealgorithm: Agentic Repository Transformation.}
\label{alg:repo-transformation}
\vspace{2pt}\hrule height 0.4pt
\smallskip
\small
\begin{algorithmic}[1]
\Require Source repository $R_s$, agent $A$
\Ensure Verified rewrite $R_q$

\LineComment{Decompose the source into behavioral module specifications.}
\State $(\{S_i\}_{i=1}^{m}, E, G) \gets A.\Call{Decompose}{R_s}$

\LineComment{Hide the source, restructure specifications, and rewrite modules.}
\State $\Call{HideSource}{R_s}$
\State $(\{T_i\}_{i=1}^{n}, E^{\prime}) \gets A.\Call{Restructure}{\{S_i\}_{i=1}^{m}, E}$
\For{$i \in \Call{TopologicalOrder}{E^{\prime}}$}
    \State $I_i \gets \{\Call{Interface}{M_j}:(j,i)\in E^{\prime}\}$
    \State $M_i \gets A.\Call{Rewrite}{T_i,I_i}$
\EndFor

\LineComment{Assemble and verify the rewritten repository.}
\State $R_q \gets A.\Call{Assemble}{M_1,\ldots,M_n}$
\State $F \gets \Call{Verify}{R_q,G}$
\While{$F\neq\emptyset$}
    \State $R_q \gets A.\Call{Repair}{R_q,F}$
    \State $F \gets \Call{Verify}{R_q,G}$
\EndWhile

\State \textbf{return} $R_q$
\end{algorithmic}
\vspace{2pt}\hrule height 0.8pt
\end{wrapfloat}
\paragraph{Repository transformation.}
We define rewriting independently of any particular model or coding harness.
The agentic repository transformation is illustrated in \LinkFigure{fig:agentic-rewriting}: the source is decomposed top-down into behavioral specifications and reconstructed bottom-up into a verified repository.
The agent first decomposes the source repository into behavioral module specifications, their dependency structure, and executable checks (Line~2).
The source implementation is then hidden. The backend restructures the
specifications into target modules with a new dependency DAG and
interface contracts (Lines~4--5). Each target module is rewritten bottom-up
from its specification and the interfaces of already reconstructed dependencies
(Lines~6--9). The rewritten modules are assembled into a new repository
(Line~11), which is verified against the behavioral checks and iteratively
repaired from execution feedback until verification succeeds (Lines~12--16). \LinkAlgorithm{alg:repo-transformation} therefore fixes the
information flow: source-derived decomposition, source-hidden reconstruction,
and execution-verified repair. Implementation details, including the rewriting
backend, bounded repair recipes, and frozen artifacts, are provided in
\LinkAppendix{app:rewrite-protocol}.
\WFclear

\paragraph{\benchpair{}.}
The pairwise design contains four relations. A \emph{source--rewrite} pair is positive
by direct derivation. A \emph{rewrite--rewrite} pair is also positive because both
implementations descend from the same source. The two negative relations
separate distinct confounders. A \emph{rewrite--independent} pair preserves purpose but removes the provenance path, making it the central hard negative. A
\emph{rewrite--unrelated} pair changes both purpose and provenance, providing an easier
control for incidental similarity. Together, these relations test whether a
detector identifies derivation rather than shared functionality, generator
style, or repository category. The relation definitions are summarized in
\LinkAppendix{app:task-construction}.

\paragraph{\benchattr{}.}
Each verified rewrite is queried against a fixed candidate set containing its
true source exactly once. Source-matched independent repositories create the
hard alternatives, while additional distractors prevent the task from reducing
to a single pairwise comparison. Candidate language is matched where possible
so that language identity does not reveal the source. A detector must therefore
rank the complete set and may abstain. Attribution is credited only when the
source is retrieved and its score exceeds the source-specific independent-code
reference band. Freezing the candidate sets and negative neighborhoods keeps
the comparison identical across detectors. Complete composition, matching
rules, and validity audits appear in \LinkAppendix{app:task-construction}.

Both tasks use the same underlying provenance records. Pairwise detection
tests a specified relationship; attribution tests recovery among competing
repositories. Frozen task identifiers, candidate sets, and reference
neighborhoods make these comparisons reproducible across detectors.

\subsection{Features of \benchmark{}}
\label{sec:features}

Conventional clone and plagiarism benchmarks primarily evaluate similarity
between localized code fragments under bounded transformations. In contrast,
\benchmark{} studies whether provenance remains recoverable after coding agents
reconstruct real repositories while preserving tested behavior. This setting
gives the benchmark several distinctive properties
(\LinkAppendix{app:benchmark-comparison}).

\paragraph{Repository-scale provenance.}
The unit of evaluation is a repository rather than a function or code
fragment. Provenance evidence may be redistributed across files, interfaces,
dependencies, and architecture, requiring detectors to aggregate weak evidence
across a large codebase. The released sources span from small projects to
repositories containing millions of code tokens
(\LinkSection{app:source-stats}).

\paragraph{Behavior-constrained agentic rewriting.}
Difficulty cannot be created simply by perturbing code until similarity
disappears. Each released rewrite is reconstructed by a coding agent and must
reproduce source-derived executable behavior before entering the benchmark.
The guarantee is deliberately limited to tested behavior rather than full
semantic equivalence, but prevents detector failure from being attributed
merely to non-functional transformations.

\paragraph{Provenance-controlled hard negatives.}
Similarity alone is insufficient evidence of derivation. \benchmark{} compares
rewrites not only with unrelated repositories but also with independent
same-domain and same-purpose implementations. These controls preserve
functionality, vocabulary, or repository category without preserving the
source-provenance path, forcing detectors to distinguish derivation from
generic semantic relatedness.

\paragraph{Evidence-aware attribution.}
Closed-set retrieval always has a nearest candidate, even when the true source
is absent or indistinguishable from independent code. \benchmark{} therefore
pairs ranking metrics with source-specific reference bands measured from
independent repositories. This separates the ability to retrieve a relatively
similar source from the stronger ability to establish source-specific
provenance evidence and enables explicit abstention diagnostics.

\paragraph{Same- and cross-language provenance.}
The benchmark spans both same-language and cross-language rewriting axes.
Same-language instances test reconstruction without translation, whereas
cross-language instances additionally change language-specific lexical and
syntactic cues. Together they test which provenance signals survive each
setting.

\section{Experiments}
\label{sec:detectors}
\paragraph{Evaluated models.}
We evaluate a detector suite spanning four evidence families
(\LinkTable{tab:suite}): lexical and statistical repository similarities;
token clone indexes (SourcererCC~\citep{sajnani2016sourcerercc} and
MSCCD~\citep{zhu2022msccd}); structural or learned clone detectors (the
TreeCen-inspired AST surrogate~\citep{hu2023treecen},
StoneDetector~\citep{heinze2025stone}, TACC~\citep{wang2023comparison}, and
Oreo~\citep{saini2018oreo}); and repository embeddings
(CodeBERT~\citep{feng2020codebert}, sequence-only
GraphCodeBERT~\citep{guo2020graphcodebert}, and
UniXcoder~\citep{guo2022unixcoder}). These families inspect different evidence
and have different repository-level applicability limits. Implementation
provenance, language scope, checkpoints, and
repository adapters are detailed in
\LinkAppendix{app:release-suite-detail}. Rewrites use Codex CLI 0.144.5
with GPT-5.6-Sol as the default generator backbone; additional backbone
ablations use GLM-5.3 and DeepSeek-V4.1-Flash
(\LinkAppendix{app:generator-effects}). The generator backend, the test
procedure of each detector family, and our Java-specific adaptations are
summarized in \LinkAppendix{app:impl-details}.

\paragraph{Evaluation protocol.}
Every detector receives the same frozen query and candidate repositories and
is evaluated on both tasks from \LinkSection{sec:tasks}: pairwise detection
uses the $\nPair$ positive pairs and their same-domain
negative pairs, whereas attribution uses the fixed $20$-candidate sets for
$\nAttr$ queries and permits abstention. We report AUROC for pairwise
discrimination and best F1 as a retrospective threshold summary, calibrated
TPR/FPR under each source-specific independent-code band, and Rank@1 plus
calibrated attribution for source identification; native
region outputs are aggregated through fixed repository adapters, embedding
models use frozen windowing and top-$k$ pooling.
Both calibrated decisions share one source-specific independent-code band,
$\beta_d(s)=\max_{N\in\mathcal{N}_s}d(R_s,N)$, defined by the source's
measured same-domain neighborhood:
\begin{equation}
\underbrace{\hat{y}_d(q,c)=\mathbf{1}\!\left[d(R_q,R_c)>\beta_d(s)\right]}_{\text{pairwise decision}}
\qquad
\underbrace{g_{\mathrm{cal}}(q;d)=g_{\mathrm{rank}}(q;d)\,\mathbf{1}\!\left[d(R_q,R_s)>\beta_d(s)\right]}_{\text{calibrated attribution}}
\label{eq:det-protocol}
\end{equation}
with $g_{\mathrm{rank}}$ the tie-aware Rank@1 credit of
\LinkAppendix{app:task-construction}.
Calibrated attribution is source-conditional evaluation credit, not a
measured deployment abstention rate. Source-absence decisions are evaluated
separately as a diagnostic; provisional native rows lack validated
reference bands.
The complete input, adaptation, calibration, abstention, and
execution-integrity contract appears in
\LinkAppendix{app:protocol-contract}.

\subsection{Main Results}
\label{sec:main}

\begin{table*}[t]
\centering
\tablestyle
\caption{Complete static-detector suite. Common-protocol methods are evaluated
on all $\nPair$ Pair instances and $\nAttr$ Attribution queries. Methods marked
$\dagger$ do not support valid outputs on the full benchmark and are therefore
reported only as provisional native results; their source-specific reference
bands are not validated, so calibrated metrics are omitted. Calibrated
attribution requires the true source to be both top-ranked and above its
reference band.}
\label{tab:suite}
\resizebox{\textwidth}{!}{%
\begin{tabular}{ll|cccc|cc}
\toprule
\rowcolor{TableHeader}
\tablehead{Family} & \tablehead{Detector}
& \multicolumn{4}{c|}{\tablehead{Pairwise}}
& \multicolumn{2}{c}{\tablehead{Attribution}}\\
\cmidrule(lr){3-6}\cmidrule(l){7-8}
\rowcolor{TableHeader}
& & \tablehead{AUROC $\uparrow$}
& \tablehead{Best F1 $\uparrow$}
& \tablehead{Cal.\ TPR $\uparrow$}
& \tablehead{Cal.\ FPR $\downarrow$}
& \tablehead{Rank@1 $\uparrow$}
& \tablehead{Cal.\ attr.\ $\uparrow$}\\
\midrule
\multirow{4}{*}{\makecell[l]{Lexical /\\statistical}}
& Token TF--IDF Cosine & 0.493 & 0.350 & 0.044 & 0.007 & 0.050 & 0.000 \\
& Character 5-gram Similarity & 0.412 & 0.297 & 0.009 & 0.028 & 0.007 & 0.001 \\
& Repository Statistical Profile & 0.490 & 0.301 & 0.014 & 0.032 & 0.000 & 0.000 \\
& Token-set Jaccard & 0.288 & 0.297 & 0.032 & 0.128 & 0.000 & 0.000 \\
\midrule
\multirow{2}{*}{Token clone}
& SourcererCC$^\dagger$~\citep{sajnani2016sourcerercc} & 0.501 & 0.299 & --- & --- & 0.048 & --- \\
& MSCCD$^\dagger$~\citep{zhu2022msccd} & 0.536 & 0.298 & --- & --- & 0.113 & --- \\
\midrule
\multirow{4}{*}{\makecell[l]{Structural /\\learned clone}}
& AST surrogate & 0.558 & 0.319 & 0.124 & 0.128 & 0.050 & 0.010 \\
& StoneDetector$^\dagger$~\citep{heinze2025stone} & 0.632 & \textbf{0.464} & --- & --- & 0.062 & --- \\
& TACC$^\dagger$~\citep{wang2023comparison} & 0.557 & 0.434 & --- & --- & 0.312 & --- \\
& Oreo$^\dagger$~\citep{saini2018oreo} & 0.578 & 0.432 & --- & --- & 0.000 & --- \\
\midrule
\multirow{3}{*}{\makecell[l]{Repository\\embedding}}
& CodeBERT~\citep{feng2020codebert} & 0.583 & 0.318 & 0.074 & 0.034 & 0.060 & 0.010 \\
& GraphCodeBERT (sequence-only)~\citep{guo2020graphcodebert} & 0.591 & 0.320 & 0.104 & 0.058 & 0.090 & 0.020 \\
& UniXcoder~\citep{guo2022unixcoder} & \textbf{0.661} & 0.394 & 0.023 & \textbf{0.004} & \textbf{0.220} & 0.010 \\
\bottomrule
\end{tabular}
}
\end{table*}

\paragraph{Detector families retain different amounts of usable evidence.}
\LinkTable{tab:suite} separates detection performance from tool applicability.
Lexical methods provide weak discrimination: no such method
exceeds AUROC $0.493$, and their calibrated attribution is at most $0.1\%$.
Native clone tools have limited readable coverage and incomplete candidate sets. Their
provisional AUROC ranges from $0.501$ to $0.632$, but readable coverage
ranges from 60 to 368 Pair instances and from 11 to 69 Attr queries.
Embeddings cover the full benchmark, with AUROC $0.583$--$0.661$ and Rank@1
$0.060$--$0.220$, but calibrated attribution remains $0.010$--$0.020$.
\LinkAppendix{app:metrics-calibration} maps each
family to its evidence and applicability limits. The provisional native rows
do not support a full-benchmark ranking against the encoders.

\begin{figure*}[t]
\centering
\includegraphics[width=\textwidth]{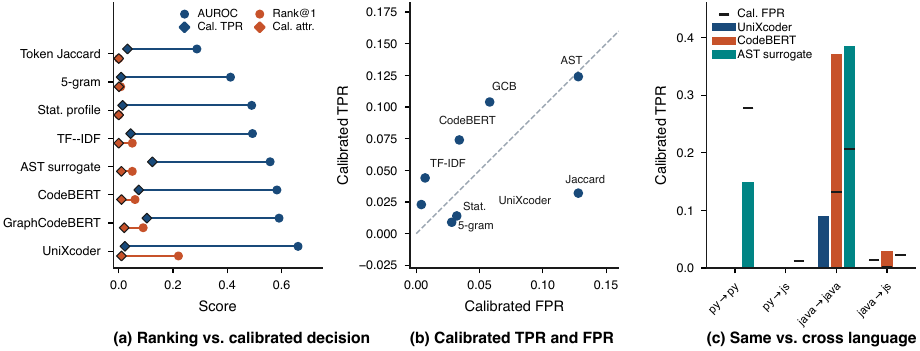}
\caption{\textbf{(a)} Discrimination and retrieval alongside calibrated
provenance decisions for the eight common-protocol detectors. Connectors
pair AUROC with calibrated TPR and Rank@1 with calibrated attribution;
AUROC and TPR measure different quantities, so their distance is not a
performance drop on one metric.
\textbf{(b)} Calibrated true- and false-positive rates per detector. The
dashed line marks equal rates. \textbf{(c)} Calibrated TPR (bars) and FPR
(ticks) on the four major language axes for three representative detectors.
Same-language Java retains signal, with limited calibrated detection.
Overall scores follow \hyperref[tab:suite]{Table~\ref*{tab:suite}};
\LinkAppendix{app:stratified-results} gives the per-axis results.}
\label{fig:main-failures}
\end{figure*}

\paragraph{Ranking is not a provenance decision.}
UniXcoder ranks the true source first for $22\%$ of queries, but only $1\%$
receive calibrated attribution credit. GraphCodeBERT similarly reaches
$9\%$ Rank@1 and $2\%$ calibrated attribution. Across the encoders, pairwise
AUROC is $0.58$--$0.66$, while calibrated TPR is $2.3$--$10.4\%$.
Thus relative source preference often fails the independent-code evidence
requirement; ranking alone would miss this distinction.
\LinkFigure{fig:main-failures}\hyperref[fig:main-failures]{a}--\hyperref[fig:main-failures]{b}
visualize the gap for every calibrated detector. See
\hyperref[tab:suite]{Table~\ref*{tab:suite}}.

\paragraph{Translation amplifies the difficulty but is not required.}
The residual signal is strongest in same-language Java: UniXcoder reaches
AUROC $0.834$, but only $9.0\%$ calibrated TPR. Detectors with higher TPR on
this axis also incur FPR as high as $24.3\%$. More accepted positives therefore
need not mean more source-specific evidence. The within-source comparison
in \LinkAppendix{app:language-pairing} also finds lower encoder margins after
cross-language rewriting, while same-language margins remain low.
\LinkFigure{fig:main-failures}\hyperref[fig:main-failures]{c} compares
calibrated detection across the four major axes.

\paragraph{Scale and input scope do not fully explain the gap.}
UniXcoder's median margin remains negative across the source-complexity
terciles. Restricting the source to oracle-anchored files changes its score by a
median $-0.033$, with few additional band passes.
These diagnostics provide no broad recovery of calibrated detection through
source-size stratification or removal of untested files
(\LinkAppendix{app:scale-scope}).

\begin{figure*}[t]
\centering
\includegraphics[width=\textwidth]{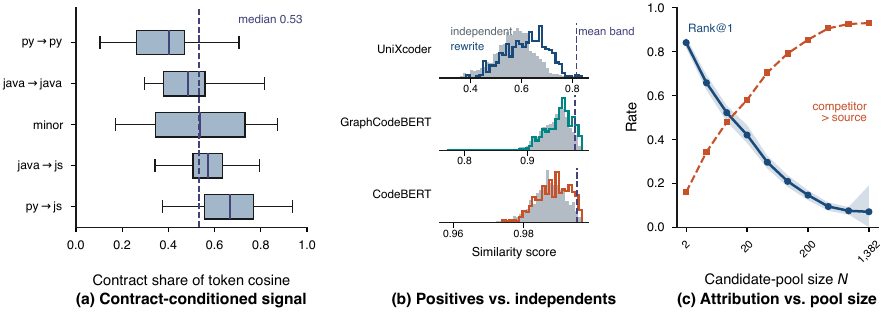}
\caption{\textbf{(a)} Share of token-frequency cosine contributed by
contract vocabulary, by language axis, with the remaining axes pooled. The
dashed line marks the benchmark
median $0.53$ in the contract-vocabulary diagnostic; IDF weighting is omitted.
\textbf{(b)} Rewrite-to-source and independent-candidate score
distributions for the three embedding detectors, with the mean per-source
band (dashed). The distributions overlap under each encoder; the mean band
is a visual reference, not a pooled decision threshold. \textbf{(c)}
Attribution Rank@1 and the rate of drawing at least one independent
competitor scoring above the true source as the candidate pool grows toward the full
$1{,}382$-repository pool in the candidate-pool sensitivity analysis. The primary task fixes 20
candidates per query.}
\label{fig:main-mechanisms}
\end{figure*}

\paragraph{Contract vocabulary explains part of the surviving lexical overlap.}
The token-frequency cosine diagnostic assigns a median share of $0.530$
to tokens found in the behavioral contract. The share varies from $0.67$
on Python-to-JavaScript to $0.40$ on Python-to-Python, and is negatively
associated with the UniXcoder margin on the four major axes ($\rho=-0.227$).
This pattern is consistent with shared functional vocabulary preserving
lexical overlap without reliably supporting a provenance decision. The
diagnostic omits IDF weighting and does not decompose the primary TF--IDF score.
\LinkFigure{fig:main-mechanisms}\hyperref[fig:main-mechanisms]{a} shows the
axis distribution.
\LinkAppendix{app:s2-contract} reports
the axis breakdown.

\paragraph{Source absence exposes false attribution.}
Removing the true source does not force every detector to abstain. CodeBERT
makes five false attributions and UniXcoder makes two across the 20 evaluated
benchmark queries. A detector can rank
candidates while lacking evidence that the source is present.
\LinkAppendix{app:source-absence-diagnostic}
separates correct, wrong, abstained, and false-attribution outcomes.

\subsection{Discussion}
\label{sec:discussion}

\paragraph{Repository provenance requires evidence beyond retrieval.}
The results distinguish the derivation relation from the evidence available
in the rewritten artifact. A true source can remain the closest candidate
without standing above independent same-domain code. The source-specific
band makes that distinction measurable, but does not establish a universal
threshold or a deployable open-set decision rule. The source-absence
study shows why retrieval must be paired with calibration and abstention
(\LinkAppendix{app:source-absence-diagnostic}).

\paragraph{Surviving similarity need not be source-specific.}
The detector families expose different limits. Lexical methods depend on
overlap that reconstruction often reduces; native clone tools additionally
require valid parsing, region mapping, and repository aggregation. Encoders
recover more relative signal, yet shared contracts and domain conventions
can also make independent implementations similar. The contract-vocabulary
and candidate-pool diagnostics help explain why improving ranking alone
may leave the provenance decision unresolved
(\LinkFigure{fig:main-mechanisms}). These results motivate source-specific invariants alongside stronger
representations, with calibration testing whether their signal also occurs
in independent code.

\begin{figure*}[t]
\centering
\includegraphics[width=\textwidth]{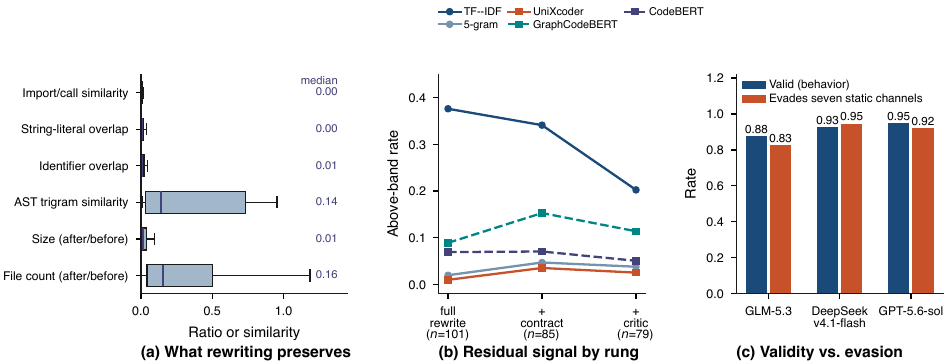}
\caption{\textbf{(a)} Source-to-artifact descriptors across 1,156 recorded
pairs: file-count and size ratios, AST-trigram cosine, and identifier,
string-literal, and import/call similarities. AST values use 400 complete
parses; medians are annotated. \textbf{(b)} Fraction of
fully conforming rewrites above the source band, by rung of the rewriting
pipeline ($n=101,85,79$). These are separate accepted cohorts, not matched
within-artifact changes. \textbf{(c)} Behavioral validity
and provenance evasion among valid artifacts for three frontier generators
under the same pipeline on 40 Python sources. Evasion refers to the seven
static channels in the generator study, not the full primary suite.}
\label{fig:main-agentic}
\end{figure*}

\paragraph{Source-hidden reconstruction constrains behavior more than implementation.}
Execution fixes the tested inputs and outputs while leaving many internal
choices open. The descriptor study records low source-to-artifact size and
identifier overlap, alongside changes in file counts and structure
(\LinkFigure{fig:main-agentic}\hyperref[fig:main-agentic]{a}). Across the
fully conforming rung cohorts, embedding band passage is already rare and
lexical band passage is lower at the critic rung
(\LinkFigure{fig:main-agentic}\hyperref[fig:main-agentic]{b}). This pattern
supports an information-bottleneck intuition: source-hidden reconstruction
can retain behavior while discarding implementation-contingent details.
It is neither an information-theoretic result nor evidence that every rewrite
changes every structural feature; the rung cohorts also differ in membership.

\paragraph{Concrete behavioral contracts matter for reconstruction.}
In the 29-source generation study, purpose-only descriptions yield no
rewrite passing the $0.99$ conformance gate, whereas the contract-only arm
reaches full conformance on $24/29$ sources. The comparison concerns these
specific recipes, but suggests that intended functionality alone
underspecifies the tested behavior. \LinkAppendix{app:case-failures} and
\LinkAppendix{app:case-llm-failure} provide the designs and outcomes.

\paragraph{Behavior preservation limits simple surface edits.}
Across 20 artifacts, the tested identifier and string-literal edits all
break behavioral checks. Comment laundering preserves behavior in $16/20$
but changes TF--IDF and $k$-gram margins by at most $0.009$. The tested
interventions thus either break behavior or barely change the measured
lexical evidence. \LinkAppendix{app:case-llm-failure}
reports the complete intervention results.

\paragraph{The generator study bounds, rather than establishes, generality.}
The default GPT-5.6-Sol backbone and two alternative backbones produce
valid rewrites that frequently fall
below the seven evaluated static channels
(\LinkFigure{fig:main-agentic}\hyperref[fig:main-agentic]{c}). The pattern is
therefore not confined to one generator in this study. The static suite's
failures do not imply that no detector can recover provenance,
and the experiment does not isolate model capability as a causal factor.

\paragraph{The benchmark supports bounded provenance claims.}
Behavioral validity concerns the tested interfaces and inputs, and the
same-domain references are screened rather than legally certified as
independent. Native clone records remain provisional, and the primary
calibrated-attribution metric is source-conditional credit. These boundaries
motivate stronger hidden behavioral tests and validated repository adapters,
with source-held-out calibration and explicit source-absence evaluation.
The unseen-input audit (\LinkAppendix{app:functional-preservation}) and detector protocol
(\LinkAppendix{app:release-suite-detail}) specify what is currently supported.

\subsection{Case Study}
\label{sec:case-study}

\paragraph{Same-language Java remains difficult.}
The Java-to-Java slice removes language translation as an explanation.
Of its Pair instances, $71\%$ fall below every construction-channel
band. TF--IDF accepts $24\%$, while UniXcoder accepts $9\%$. The failure therefore
persists even when source and artifact share a language. The per-axis records
are reported in \LinkAppendix{app:stratified-results}.

\paragraph{More accepted positives also produce more false evidence.}
On this slice, CodeBERT and GraphCodeBERT recover $37\%$ and $50\%$ of
positives. They
also accept 13.2\% and 24.3\% of the same-domain negative pairs. The apparent
gain is therefore not a clean recovery of source-specific provenance. The
complete detector records are in
\LinkAppendix{app:release-suite-detail}.

\paragraph{Source removal exposes the decision error.}
Among the 15 Java attribution queries in the source-removal diagnostic,
CodeBERT falsely names a remaining candidate five times. UniXcoder does so
once. These errors show why ranking the available candidates is insufficient
for provenance. \LinkAppendix{app:source-absence-diagnostic}
reports the complete case-study outcomes.

\section{Related Work}
\label{sec:related}

\paragraph{Code laundering.} Detectors have long been stress-tested against
transformations that hide reuse. Examples include obfuscation in software
forensics and malware lineage~\citep{cesare2013control,cheers2020plagiarism}.
Other work crafts semantics-preserving perturbations against neural code
models~\citep{yefet2020adversarial,zhang2023challenging}. Those evaluations use
fragment-level benchmarks with mined or bounded synthetic edits
\citep{svajlenko2014towards,svajlenko2015evaluating,svajlenko2016bigcloneeval,duncan2021pyclone,zhu2024development,burock2025clorero,wang2023comparison}.
Their ground truth is least reliable for semantic clones
\citep{krinke2022bigclonebench,krinke2025misuse,roy2018benchmarks}.
In \benchmark{}, the unit is the repository. Hard positives pass an execution criterion. 
\LinkAppendix{app:benchmark-comparison} gives the detailed benchmark-level comparison and makes explicit which of these requirements are absent from prior evaluation settings.

\paragraph{Code plagiarism detection.} Relevant methods include
token fingerprints~\citep{aiken1994moss,prechelt2002finding,wang2018ccaligner},
tree and graph matching~\citep{baxter1998clone,jiang2007deckard,krinke2001identifying,komondoor2001using,liu2006gplag},
clone indexes~\citep{cordy2011nicad,sajnani2016sourcerercc,saini2018oreo}, and
pre-trained code encoders~\citep{feng2020codebert,guo2020graphcodebert,guo2022unixcoder,wang2021codet5,tao2022c4,yu2019neural,buch2019learning}.
Stylometric attribution supplies another source of evidence
\citep{frantzeskou2010source,caliskan2015anonymizing,abuhamad2018large}.
All depend on some measurable lexical, structural, or stylistic trace. Neural
clone detectors already generalize poorly to unseen functionality at fragment
scale~\citep{liu2021can,pinku2024on}. Our tasks ask whether surviving relative
similarity supports a calibrated claim at repository scale.
SourceTracker/HST~\citep{gurioli2026sourcetracker} retrieves source snippets
for verbatim or renamed fragments. It uses Recall@N and MRR. Those metrics do
not test source-specific reference-band decisions against same-domain repositories. LC$^3$~\citep{lan2026lc3} extends semantic clone detection to long and cross-language code but still formulates the problem as
clone matching rather than repository-level source provenance.


\paragraph{Agentic coding for repository transformation.}
Coding agents have progressed from function-level generation
\citep{chen2021evaluating,li2022competition} to modifying real repositories
\citep{jimenez2024swebench,yang2024sweagent}. Recent benchmarks evaluate
repository transformation itself: RepoTransBench~\citep{wang2026repotransbench}
studies cross-language repository translation with executable tests, while
RepoMod-Bench~\citep{li2026repomod} evaluates repository modernization with
implementation-independent behavioral tests. These benchmarks ask whether an
agent can produce a functionally valid transformed repository. Our question
begins after transformation: whether provenance remains detectable after
agentic rewriting preserves tested behavior.

\section{Conclusion}

Coding agents make repository-scale rewriting a practical provenance problem: an implementation can be extensively reconstructed while retaining tested behavior and a real derivation path to its source. We introduced \benchmark{} to make this setting measurable, combining execution-verified agentic rewrites with independent same-domain controls and complementary Pair and Attribution tasks. Across the evaluated detector families, stronger representations recover some relative similarity, yet this signal rarely becomes source-specific evidence sufficient for calibrated attribution. The resulting ranking--decision gap is the central finding of our study: identifying the closest repository is not the same as establishing where code came from.
\benchmark{} therefore shifts repository provenance evaluation from similarity alone toward evidence-aware detection with calibration and abstention. As coding agents increasingly transform existing software rather than generate isolated snippets, reliable provenance will require methods that recover source-specific invariants surviving such reconstruction. We release the frozen artifacts and evaluation protocol to provide a reproducible foundation
for developing and testing those methods.

\clearpage
\section*{AI Use Statement}
We used generative AI to polish the manuscript and check its clarity and
consistency. Coding agents were the primary tools for collecting and
generating benchmark artifacts, including repository analysis, source-hidden
rewriting, and execution-guided repair. AI tools also assisted with
implementing and debugging experimental code, preparing figures, and checking
result interpretations. The agent configuration, generator backbones, and
prompts are documented in \LinkAppendix{app:rewrite-protocol},
\LinkAppendix{app:generator-effects}, and \LinkAppendix{app:prompt}.
Generated rewrites underwent the execution-based verification described in
\LinkAppendix{app:verification}. The authors take responsibility for the
final manuscript, experimental results, and released artifacts, including
content produced with AI assistance.

\section*{Ethics Statement}
This work studies agentic code laundering for scientific evaluation of
repository provenance and to support the development of stronger detection
methods. We acknowledge that rewriting methods could also be misused to
conceal copying or evade provenance checks. Our benchmark retains source
identities and derivation records, and documents source license metadata
(\LinkAppendix{app:source-selection} and \LinkAppendix{app:licenses}). We
intend the benchmark and pipeline for research on provenance detection;
low detector scores do not establish independent authorship or justify
disregarding source licenses. The reported provenance relationships are
technical records of derivation, not legal determinations of ownership
or infringement.

\section*{Reproducibility Statement}
We will release the benchmark data and the complete pipeline code for data
collection, agentic rewriting, behavioral verification, task construction,
and detector evaluation. \LinkSection{sec:benchmark} defines the tasks and
construction protocol. \LinkAppendix{app:construction} documents source
selection, rewriting configurations, prompts, and acceptance checks;
\LinkAppendix{app:impl-details} and \LinkAppendix{app:protocol-contract}
describe implementation details and repository-level detector adaptations.
\LinkAppendix{app:metrics-calibration} specifies the metrics, calibration,
and statistical units. These materials, together with the frozen task
manifests and evaluation records, support reproduction of the reported
comparisons.

\bibliography{iclr2027_conference}
\bibliographystyle{iclr2027_conference}

\clearpage
\appendix

\begingroup
\hypersetup{linkcolor=darkcerulean,linkbordercolor=green}%
\section*{Appendix Contents}
\label{app:outline}
\small
\newcommand{\apptocA}[2]{\par\medskip\noindent\hyperref[#1]{\bfseries \ref*{#1}\hspace{0.8em}#2\ \dotfill\ \pageref*{#1}}\par}
\newcommand{\apptocB}[2]{\par\noindent\hspace*{1.8em}\hyperref[#1]{\ref*{#1}\hspace{0.8em}#2\ \dotfill\ \pageref*{#1}}\par}
\apptocA{app:construction}{Benchmark Construction}
\apptocB{app:source-selection}{Source selection and provenance}
\apptocB{app:rewrite-protocol}{Agentic rewriting protocol}
\apptocB{app:verification}{Behavioral verification and acceptance}
\apptocB{app:task-construction}{Pair and Attribution Construction}
\apptocB{app:prompt}{Prompts}
\apptocA{app:characterization}{Benchmark Characterization}
\apptocB{app:benchmark-comparison}{Comparison with existing evaluations}
\apptocB{app:source-stats}{Source repository statistics}
\apptocB{app:functional-preservation}{Functional preservation}
\apptocB{app:rewrite-magnitude}{Rewrite magnitude and structural change}
\apptocB{app:independent-controls}{Information-access controls}
\apptocA{app:detector-evaluation}{Detector Evaluation}
\apptocB{app:impl-details}{Implementation details}
\apptocB{app:baselines}{Baselines and repository-level adaptation}
\apptocB{app:pair-results}{Complete Pair results}
\apptocB{app:attr-results}{Complete Attribution results}
\apptocB{app:java-tier}{Language-specific / Java-tier results}
\apptocA{app:analyses}{Explanatory Analyses}
\apptocB{app:surviving-signals}{What provenance signals survive rewriting?}
\apptocB{app:rewrite-evidence}{Structural and rung-wise rewrite evidence}
\apptocB{app:strength-difficulty}{Rewrite strength and detection difficulty}
\apptocB{app:language-pairing}{Same-language vs.\ cross-language provenance}
\apptocB{app:scale-scope}{Repository scale and rewrite scope}
\apptocB{app:generator-effects}{Generator-backbone ablation}
\apptocB{app:multisource}{Multi-source provenance}
\apptocA{app:case-studies}{Case Studies}
\apptocB{app:case-backend}{The complete agentic rewriting backend}
\apptocB{app:case-success}{What a successful evasion looks like}
\apptocB{app:case-clone-failure}{When lexical evidence fails}
\apptocB{app:case-llm-failure}{When an LLM / coding agent fails}
\apptocB{app:case-multisource}{Multi-source provenance}
\apptocB{app:case-failures}{Failure cases of benchmark construction}
\apptocA{app:robustness}{Robustness and Limitations}
\apptocB{app:metrics-calibration}{Metrics, calibration, and statistical units}
\apptocB{app:pool-sensitivity}{Negative/candidate-pool sensitivity}
\apptocB{app:validity-audits}{Additional validity audits}
\apptocB{app:limitations}{Limitations}
\endgroup
\clearpage

\section{Benchmark Construction}
\label{app:construction}

This part documents how the benchmark was constructed:
\LinkSection{app:source-selection} the sources and their provenance,
\LinkSection{app:rewrite-protocol} the frozen agentic rewriting protocol,
\LinkSection{app:verification} behavioral verification and acceptance, and
\LinkSection{app:task-construction} the deterministic construction of the two
tasks. \LinkSection{app:prompt} reproduces the concrete stage prompts.
\benchmark{} combines repository-level provenance tasks with agentic
rewriting and execution-based acceptance. The comparison in
\LinkTable{tab:compare} concerns this combination, rather than priority over
repository transformation benchmarks in general.

\paragraph{Release policy: the full verified set.}
The published benchmark is the full frozen collection of
construction-verified instances. Every released instance passed the same
construction gates, and no detector-informed or difficulty-based selection
defines the release: the ``challenge tier'' referenced in later
analyses is a descriptive stratum of the released instances (below every
construction band with substantive size), not a separately selected release
set. A random control tier is released alongside the main instances so that
selection-conditioned claims can be checked against unselected pairs.

\paragraph{Functional verification of every rewrite.}
A rewrite enters the benchmark only after it is verified to preserve its
source's tested behavior to the stated acceptance threshold. Verification is execution-based, not
model-judged. A language-agnostic conformance harness first captures golden
input--output traces by executing the \emph{original} repository on a case
list covering its core behavior; the rewrite is then executed on the same
cases through a thin adapter that is excluded from all scoring. The release
gates require the source harness to yield at least five successful cases,
and the rewrite to reproduce at least half of the expected outputs exactly
with overall conformance of at least $0.5$; agreement on rejected inputs
alone is insufficient. Two further validity gates reject degenerate evasion:
the adapter must stay thin (at most half of the artifact's tokens), and
artifact size is judged against $\min(\text{original tokens},
\refOrigTokens)$. This capped denominator limits degenerate small artifacts
without requiring whole-source size retention. Of the scored candidates, \nOraclePassed{} pass the
behavioral oracle and \nVerified{} remain after the adapter check;
\LinkSection{app:verification} reports the acceptance audits, including
exact-match rates, an unseen-input audit, and a watermark scan of the
released artifacts.

\subsection{Source selection and provenance}
\label{app:source-selection}
\paragraph{Dataset units and provenance.}
The published benchmark contains $\nSources$ pinned source repositories:
$\nSourcesPy$ Python and $\nSourcesJava$ Java. Every source contributes
$\nPairPerSource$ \benchpair{} instances, so the pairwise task spans
$\nSources\times\nPairPerSource=\nPairRelease$ verified pairs. \benchattr{}
queries are anchored on the sources that carry a full set of five measured
same-domain reference repositories, because each attribution candidate set
embeds the true source among these neighbors as hard negatives; $\nAttrEligible$
of the $\nSources$ sources satisfy this condition. Each attribution query
contains 20 fixed candidates and the true source exactly once. The benchmark
also contains the measured same-domain reference neighborhoods needed to
reproduce every source-specific band. These neighborhoods are empirical
controls. They are not certificates of legal or historical independence.

\paragraph{Source repository roster.}
The released manifest pins every source repository's URL, commit, and snapshot
date. The roster below lists all $\nSources$ sources, grouped by dominant
code language (see \LinkFigure{fig:repo-stats} for their composition and
scale). Names in {\color{darkcerulean}blue} mark the $\nAttrEligible$ sources that
carry five measured same-domain reference repositories and therefore anchor
\benchattr{} attribution queries.

{\scriptsize\raggedright
\par\noindent\textbf{Python (199):} \textcolor{darkcerulean}{3ES}, \textcolor{darkcerulean}{ACES}, \textcolor{darkcerulean}{AirbrakesV2}, \textcolor{darkcerulean}{AlexaPi}, \textcolor{darkcerulean}{Annif}, \textcolor{darkcerulean}{AutoGPT}, \textcolor{darkcerulean}{ChatGLM-6B}, \textcolor{darkcerulean}{ComfyUI}, \textcolor{darkcerulean}{Deep-Live-Cam}, \textcolor{darkcerulean}{DeepSeek-V3}, \textcolor{darkcerulean}{Fooocus}, GitPython, GreedyBear, \textcolor{darkcerulean}{LLaMA-Factory}, \textcolor{darkcerulean}{MONAI}, \textcolor{darkcerulean}{MSS}, \textcolor{darkcerulean}{MetaGPT}, \textcolor{darkcerulean}{NeMo}, \textcolor{darkcerulean}{OpenHands}, \textcolor{darkcerulean}{PaddleOCR}, PlasmaPy, \textcolor{darkcerulean}{Real-Time-Voice-Cloning}, \textcolor{darkcerulean}{TLOmodel}, \textcolor{darkcerulean}{TagStudio}, \textcolor{darkcerulean}{TransformerLens}, \textcolor{darkcerulean}{VDRS-Solutions}, \textcolor{darkcerulean}{accelerate}, \textcolor{darkcerulean}{agenta}, \textcolor{darkcerulean}{aiohttp}, \textcolor{darkcerulean}{aiortc}, \textcolor{darkcerulean}{albumentations}, \textcolor{darkcerulean}{ansible}, \textcolor{darkcerulean}{ansible-cartridge}, app, arkouda, \textcolor{darkcerulean}{backend}, \textcolor{darkcerulean}{certbot}, \textcolor{darkcerulean}{chainlit}, \textcolor{darkcerulean}{checkov}, ckan, \textcolor{darkcerulean}{conan}, \textcolor{darkcerulean}{core}, \textcolor{darkcerulean}{cpython}, \textcolor{darkcerulean}{crosstl}, \textcolor{darkcerulean}{cupy}, \textcolor{darkcerulean}{cython}, \textcolor{darkcerulean}{dask}, \textcolor{darkcerulean}{dask-cuda}, \textcolor{darkcerulean}{dgl}, \textcolor{darkcerulean}{diffusers}, \textcolor{darkcerulean}{django}, django-celery-beat, \textcolor{darkcerulean}{django-ninja}, \textcolor{darkcerulean}{django-two-factor-auth}, \textcolor{darkcerulean}{docling}, \textcolor{darkcerulean}{dspy}, dvc, \textcolor{darkcerulean}{face\_recognition}, \textcolor{darkcerulean}{faceswap}, fapolicy-analyzer, \textcolor{darkcerulean}{fastapi}, faster-whisper, \textcolor{darkcerulean}{feast}, \textcolor{darkcerulean}{flask}, \textcolor{darkcerulean}{flet}, \textcolor{darkcerulean}{fractal-tasks-core}, \textcolor{darkcerulean}{gorilla}, \textcolor{darkcerulean}{gpt-engineer}, \textcolor{darkcerulean}{gpt4free}, \textcolor{darkcerulean}{gpt\_academic}, \textcolor{darkcerulean}{grok-1}, \textcolor{darkcerulean}{hackingtool}, \textcolor{darkcerulean}{hal-cgp}, \textcolor{darkcerulean}{imod-python}, \textcolor{darkcerulean}{impacket}, \textcolor{darkcerulean}{instructor}, \textcolor{darkcerulean}{ivadomed}, jax, \textcolor{darkcerulean}{joinmarket-clientserver}, jwcrypto, \textcolor{darkcerulean}{kedro}, \textcolor{darkcerulean}{keras}, klein, \textcolor{darkcerulean}{knowledge-repo}, \textcolor{darkcerulean}{kornia}, \textcolor{darkcerulean}{langflow}, \textcolor{darkcerulean}{langgraph}, \textcolor{darkcerulean}{litellm}, \textcolor{darkcerulean}{llama}, \textcolor{darkcerulean}{localstack}, \textcolor{darkcerulean}{locust}, \textcolor{darkcerulean}{loguru}, \textcolor{darkcerulean}{luigi}, manage.get.gov, manim, \textcolor{darkcerulean}{matchms-backup}, \textcolor{darkcerulean}{mathesar}, \textcolor{darkcerulean}{matplotlib}, \textcolor{darkcerulean}{meson}, \textcolor{darkcerulean}{metaflow}, \textcolor{darkcerulean}{micropython-lib}, microsoft-authentication-library-for-python, \textcolor{darkcerulean}{mlflow}, \textcolor{darkcerulean}{modin}, \textcolor{darkcerulean}{moto}, \textcolor{darkcerulean}{moviepy}, \textcolor{darkcerulean}{mypy}, \textcolor{darkcerulean}{nautobot}, \textcolor{darkcerulean}{netbox}, \textcolor{darkcerulean}{networkx}, \textcolor{darkcerulean}{nltk}, \textcolor{darkcerulean}{numba}, \textcolor{darkcerulean}{numpy}, oauthenticator, \textcolor{darkcerulean}{odoo}, \textcolor{darkcerulean}{okta-jwt-verifier-python}, open-MaStR, \textcolor{darkcerulean}{open-interpreter}, \textcolor{darkcerulean}{openlibrary}, openllmetry, openwisp-users, \textcolor{darkcerulean}{optimum-benchmark}, \textcolor{darkcerulean}{optuna}, \textcolor{darkcerulean}{paf}, \textcolor{darkcerulean}{pandas}, \textcolor{darkcerulean}{paperless-ngx}, \textcolor{darkcerulean}{phidata}, \textcolor{darkcerulean}{pip}, \textcolor{darkcerulean}{plone.restapi}, \textcolor{darkcerulean}{prefect}, \textcolor{darkcerulean}{private-gpt}, \textcolor{darkcerulean}{prowler}, \textcolor{darkcerulean}{prymer}, \textcolor{darkcerulean}{pulp\_rpm}, \textcolor{darkcerulean}{pydantic}, \textcolor{darkcerulean}{pylint}, pymapdl, \textcolor{darkcerulean}{pyopenssl}, \textcolor{darkcerulean}{pypdf}, \textcolor{darkcerulean}{pytorch}, \textcolor{darkcerulean}{pytorch-lightning}, \textcolor{darkcerulean}{qiskit}, \textcolor{darkcerulean}{quiz-backend}, \textcolor{darkcerulean}{ragas}, \textcolor{darkcerulean}{raster-vision}, \textcolor{darkcerulean}{ray}, \textcolor{darkcerulean}{real-intent}, \textcolor{darkcerulean}{repo\_standards\_validator}, \textcolor{darkcerulean}{requests}, \textcolor{darkcerulean}{rich}, \textcolor{darkcerulean}{robotframework}, \textcolor{darkcerulean}{roms-tools}, rq, \textcolor{darkcerulean}{rucio}, \textcolor{darkcerulean}{sancus-compiler}, \textcolor{darkcerulean}{scapy}, \textcolor{darkcerulean}{scikit-learn}, scipy, \textcolor{darkcerulean}{scrapy}, \textcolor{darkcerulean}{screenshot-to-code}, \textcolor{darkcerulean}{securedrop-client}, \textcolor{darkcerulean}{sentence-transformers}, sgkit, \textcolor{darkcerulean}{sherlock}, \textcolor{darkcerulean}{simtools}, \textcolor{darkcerulean}{sisyphus}, \textcolor{darkcerulean}{sktime}, \textcolor{darkcerulean}{sopel}, \textcolor{darkcerulean}{speechbrain}, \textcolor{darkcerulean}{sphinx}, \textcolor{darkcerulean}{sqlfluff}, \textcolor{darkcerulean}{sqlglot}, \textcolor{darkcerulean}{stable-diffusion-webui}, stcal, \textcolor{darkcerulean}{streamlit}, \textcolor{darkcerulean}{supervision}, \textcolor{darkcerulean}{sympy}, synapse, tardis, \textcolor{darkcerulean}{tech-blog-hub-site}, \textcolor{darkcerulean}{text-generation-webui}, \textcolor{darkcerulean}{textec}, \textcolor{darkcerulean}{tfl-live-tracker}, \textcolor{darkcerulean}{thefuck}, \textcolor{darkcerulean}{transformers}, \textcolor{darkcerulean}{trl}, \textcolor{darkcerulean}{ultralytics}, \textcolor{darkcerulean}{uxarray}, \textcolor{darkcerulean}{viztracer}, \textcolor{darkcerulean}{vllm}, xcdat, \textcolor{darkcerulean}{xclim}, \textcolor{darkcerulean}{xformers}, \textcolor{darkcerulean}{yfinance}, yolov5, \textcolor{darkcerulean}{youtube-dl}, \textcolor{darkcerulean}{yt-dlp}, zarr-python, zulip\par
\smallskip
\par\noindent\textbf{Java (83):} JSONassert, JSqlParser, \textcolor{darkcerulean}{JavaHamcrest}, \textcolor{darkcerulean}{JsonPath}, \textcolor{darkcerulean}{alluxio}, \textcolor{darkcerulean}{antlr4}, \textcolor{darkcerulean}{assertj}, avro, \textcolor{darkcerulean}{calcite}, \textcolor{darkcerulean}{cassandra}, \textcolor{darkcerulean}{checkstyle}, \textcolor{darkcerulean}{commonmark-java}, \textcolor{darkcerulean}{commons-beanutils}, \textcolor{darkcerulean}{commons-cli}, commons-codec, \textcolor{darkcerulean}{commons-collections}, \textcolor{darkcerulean}{commons-compress}, \textcolor{darkcerulean}{commons-configuration}, \textcolor{darkcerulean}{commons-csv}, \textcolor{darkcerulean}{commons-dbutils}, commons-digester, commons-geometry, \textcolor{darkcerulean}{commons-imaging}, \textcolor{darkcerulean}{commons-io}, commons-jexl, commons-jxpath, \textcolor{darkcerulean}{commons-lang}, commons-math, commons-numbers, \textcolor{darkcerulean}{commons-pool}, commons-rng, \textcolor{darkcerulean}{commons-statistics}, \textcolor{darkcerulean}{commons-text}, \textcolor{darkcerulean}{commons-validator}, doris, \textcolor{darkcerulean}{dubbo}, \textcolor{darkcerulean}{elasticsearch}, fastjson2, flexmark-java, \textcolor{darkcerulean}{flink}, \textcolor{darkcerulean}{google-java-format}, \textcolor{darkcerulean}{graphql-java}, \textcolor{darkcerulean}{grpc-java}, \textcolor{darkcerulean}{gson}, \textcolor{darkcerulean}{h2database}, \textcolor{darkcerulean}{hadoop}, \textcolor{darkcerulean}{jackson-core}, \textcolor{darkcerulean}{jackson-databind}, java-diff-utils, \textcolor{darkcerulean}{java-jwt}, \textcolor{darkcerulean}{javaparser}, \textcolor{darkcerulean}{jcommander}, jgrapht, jjwt, \textcolor{darkcerulean}{jline3}, joda-money, \textcolor{darkcerulean}{joda-time}, \textcolor{darkcerulean}{json-schema-validator}, \textcolor{darkcerulean}{jsoup}, jts, \textcolor{darkcerulean}{junit5}, \textcolor{darkcerulean}{kafka}, \textcolor{darkcerulean}{minimal-json}, \textcolor{darkcerulean}{mockito}, \textcolor{darkcerulean}{msgpack-java}, \textcolor{darkcerulean}{mvel}, \textcolor{darkcerulean}{nacos}, \textcolor{darkcerulean}{netty}, \textcolor{darkcerulean}{opensearch}, picocli, \textcolor{darkcerulean}{pinot}, \textcolor{darkcerulean}{rocketmq}, \textcolor{darkcerulean}{sentinel}, \textcolor{darkcerulean}{shardingsphere}, \textcolor{darkcerulean}{snakeyaml}, \textcolor{darkcerulean}{sonarqube}, \textcolor{darkcerulean}{spotbugs}, \textcolor{darkcerulean}{spring-boot}, \textcolor{darkcerulean}{spring-framework}, threeten-extra, \textcolor{darkcerulean}{typetools}, univocity-parsers, \textcolor{darkcerulean}{zxing}\par
}

\paragraph{Released artifacts.}
We release the rewritten repositories: for every accepted artifact, the
rewritten project, its excluded conformance adapter, the oracle traces, the
candidate lists, and the detector inputs, together with the pinned source
manifest and the per-instance task definitions of both tasks.

\paragraph{Licenses and the released pipeline.}
The sources are third-party projects with the archived license metadata
summarized in \LinkTable{tab:licenses}. The benchmark records technical
derivation; it does not determine legal ownership or license compliance.
To make the transformation process auditable, we also release the
complete data pipeline: the agentic rewriting backend and its prompts
(\LinkSection{app:rewrite-protocol} and \LinkSection{app:prompt}), the
oracle-generation and conformance harness, the verification and selection
gates (\LinkSection{app:verification}), and the freeze manifests, so the
benchmark can be audited, regenerated, and extended under the same protocol.

\subsection{Agentic rewriting protocol}
\label{app:rewrite-protocol}
\label{app:release-config}
\begin{table}[t]
\centering
\caption{Codex CLI 0.144.5 configuration with the default GPT-5.6-Sol backbone. Budgets are
configured limits, not measured costs.}
\label{tab:release-config-detail}
\tablestyle\footnotesize
\begin{tabularx}{\linewidth}{lLL}
\toprule
\rowcolor{TableHeader}
\tablehead{Stage} & \tablehead{Contract / operation} & \tablehead{Target / budget} \\
\midrule
Full contract & Extracted behavior with statistical camouflage & JS, Python, Java; $\leq3$ attempts \\
External vocabulary & Contract vocabulary shared across implementations & JS, Python, Java; $\leq3$ attempts \\
Critic & Feed surviving source names back to the agent & $\leq2$ rounds $\times$ $\leq2$ attempts \\
Size repair & Expand the implementation against its specification & JS only; $\leq3$ attempts \\
\bottomrule
\end{tabularx}
\end{table}

\LinkTable{tab:release-config-detail} summarizes the Codex backend. The agent
reads repository context, extracts a behavioral contract, constructs a new
project with the source hidden, materializes the files, and executes the
candidate. Execution failures and leaked identifiers are returned to the
agent for bounded repair. The same backend is used across all language axes.
The benchmark freezes the resulting repositories, adapters, oracle traces,
candidate lists, and detector inputs.

\subsection{Behavioral verification and acceptance}
\label{app:verification}
\paragraph{Behavioral audit.}
Behavioral validity is oracle-covered rather than whole-program equivalence.
On generation-visible successful cases, $62\%$ of Pair artifacts
reproduce every observed output exactly. The bounded benchmark unseen
input audit gives 60/63 exact matches among inputs on which the original
succeeds. By major axis, the counts are 14/14, 13/15, 16/17, and 17/17. These
checks bound the claim to the tested interfaces and inputs.

\paragraph{Observed safety behavior and watermark audit.}
The research-context prompts describe a provenance benchmark and do not ask
the agent to remove a watermark. Every Pair artifact contains code
rather than a recorded policy refusal. We searched the benchmark artifacts,
excluding dependencies and
oracle files, for case-insensitive explicit signatures including
\texttt{AI-generated}, \texttt{generated by OpenAI/ChatGPT/GPT/Codex},
\texttt{provenance watermark}, and \texttt{watermarked by}; the audit returns
zero matches. This is an output-level observation, not a claim about unlogged
failed calls, imperceptible watermarks, or other providers. Its security
implication is narrow but direct: the evaluated coding-agent workflow neither
refused the observed rewriting requests nor volunteered a visible provenance
signal on the scored outputs.

\subsection{Pair and Attribution Construction}
\label{app:task-construction}
\label{app:datasets}

\LinkTable{tab:bench-stats} summarizes the composition of the two tasks.

\paragraph{Provenance paths, not surface similarity, define the labels.}
Let $R$ be a source repository. In the positive branch, an agent first derives
a behavioral specification $D_R$ from $R$ and then produces a rewrite $R'$:
\begin{equation}
R \longrightarrow D_R \longrightarrow R'.
\end{equation}
Hiding the source implementation after extraction does not make $R'$ independent:
$D_R$ is a source-derived representation, so the provenance path remains. Two
separately generated rewrites $R'_1$ and $R'_2$ are likewise a positive pair
because both descend from $R$.

The hard-negative branch requires a genuinely source-independent description
$D_{\perp}$, such as a public task description or manually normalized
functional requirements prepared without access to $R$ or any representation
extracted from it. Generation then follows
\begin{equation}
D_{\perp} \longrightarrow I,
\qquad R \not\longrightarrow D_{\perp},
\end{equation}
and $(R',I)$ is negative despite matched purpose. By contrast, generating $I$
from a sketch extracted from $R$ would create the path
$R\rightarrow D_R\rightarrow I$ and must be labeled positive. This distinction
prevents source hiding from being mistaken for provenance independence.

\begin{table}[h!]
\centering
\caption{Four conceptual \benchpair{} relations. The hard negative preserves
functional relatedness while removing the source-provenance path.}
\label{tab:app-pair-relations}
\tablestyle\footnotesize
\begin{tabularx}{\linewidth}{LLLL}
\toprule
\rowcolor{TableHeader}
\tablehead{Relation} & \tablehead{Pair} & \tablehead{Label} & \tablehead{Meaning} \\
\midrule
Source--rewrite & $(R,R'_1)$ & $+$ & direct provenance \\
Rewrite--rewrite & $(R'_1,R'_2)$ & $+$ & shared provenance through $R$ \\
Rewrite--independent & $(R',I)$ & $-$ & matched purpose without a path from $R$ \\
Rewrite--unrelated & $(R',N)$ & $-$ & neither matched purpose nor provenance \\
\bottomrule
\end{tabularx}
\end{table}

The two negative relations form a difficulty hierarchy. $(R',N)$ is an easy
negative because functionality and provenance both differ. $(R',I)$ is the
central hard negative because functionality is held similar while provenance
changes. The controlled independent-generation setting holds the coding-agent
backbone, agent harness, compute budget, target language, and requested
functionality fixed. It varies only whether the agent receives source-derived
information or a source-independent functional sketch. This construction tests
whether a detector identifies provenance rather than generator style, language,
repository category, or generic coding conventions. The reference bands use
pinned, measured same-domain repositories discovered independently of the
rewrite process.

\paragraph{\benchpair{} instantiation.}
Each released query is a verified rewrite $R'$ paired positively with its source
$R$ and negatively with measured non-source repositories. Same-domain neighbors
provide hard negatives, while unrelated repositories provide an easier control.
The neighbors are screened for copies and pinned at fixed commits, but their
independence remains an empirical dataset assumption rather than a legal or
historical certificate. The fixed release contains $\nPair$ positive queries
from $\nSources$ sources over four rewriting axes. Every query retains the
same source-specific negative neighborhood across detectors.

\paragraph{\benchattr{} instantiation.}
Each verified rewrite $R_q$ defines an attribution query over a fixed set of 20
distinct candidates $\mathcal{C}_q$ containing its true source $R_s$ exactly
once. The remaining candidates are non-source repositories. Measured
same-domain neighbors serve as hard negatives when available, and random
distractors complete the set. Candidate identities and ordering are fixed
across detectors, so every method solves the same retrieval problem.

Given detector score $d$, let
\begin{equation}
T_q(d)
:=
\arg\max_{R_c\in\mathcal{C}_q} d(R_q,R_c)
\end{equation}
denote the set of candidates tied for the highest score. Rank@1 gives the true
source full credit when it is the unique maximum and fractional credit when it
shares the maximum:
\begin{equation}
g_{\mathrm{rank}}(q;d)
=
\frac{\mathbf{1}[R_s\in T_q(d)]}{|T_q(d)|}.
\label{eq:attr-rank-gate}
\end{equation}
This is a closed-set retrieval measure: some candidate is always top-ranked,
even when no score constitutes meaningful provenance evidence.

For each source $R_s$ and detector $d$, the source's measured same-domain
neighborhood $\mathcal{N}_s$ defines the reference band
\begin{equation}
\beta_d(s)
=
\max_{N\in\mathcal{N}_s} d(R_s,N).
\label{eq:attr-band}
\end{equation}
The true source passes the evidence gate only when its query score strictly
exceeds this source-specific band, $d(R_q,R_s)>\beta_d(s)$. Calibrated
attribution therefore requires both retrieval and evidence:
\begin{equation}
g_{\mathrm{cal}}(q;d)
=
g_{\mathrm{rank}}(q;d)\,
\mathbf{1}\!\left[d(R_q,R_s)>\beta_d(s)\right].
\label{eq:calibrated-attribution}
\end{equation}
Thus Rank@1 measures whether the source is relatively closest among the
candidates, whereas calibrated attribution asks the stronger question of
whether that retrieval is also supported beyond the similarity already
observed between the source and independent same-domain repositories.

\paragraph{Abstention and source absence.}
The evidence gate permits abstention when no candidate exceeds the calibrated
threshold. To test this behavior outside the source-present setting, we remove
$R_s$ from selected candidate sets while retaining the remaining 19 candidates
and the original threshold. Predicting any remaining candidate is a false
attribution; predicting none is a correct rejection. This diagnostic is
reported separately from the primary source-present attribution score.

\begin{table}[t]
\centering
\caption{\benchmark{} composition. Pair and Attribution provide complementary
evaluation views of the same provenance records.}
\label{tab:bench-stats}
\tablestyle
\begin{tabularx}{\linewidth}{>{\hsize=.85\hsize}Lr>{\hsize=1.15\hsize}L}
\toprule
\rowcolor{TableHeader}
\tablehead{Component} & \tablehead{Size} & \tablehead{Benchmark role} \\
\midrule
Source repositories & \nSources & real projects; \nSourcesPy{} Python, \nSourcesJava{} Java \\
Same-domain negatives & \nNegatives & measured; sim $<0.35$; median $\nNegMedianSim$ \\
\rowcolor{TableFocus}
\benchpair{} & \nPair & fixed repository-pair detection task \\
\rowcolor{TableFocus}
\benchattr{} & \nAttr & queries with 20 fixed candidates each; attribution with abstention \\
\bottomrule
\end{tabularx}
\end{table}

\paragraph{Benchmark coverage.}
The benchmark spans four rewriting axes: py$\to$js, py$\to$py,
java$\to$java, and java$\to$js. The benchmark
artifact, source, candidate-list, oracle, and score identifiers are frozen so
that every reported result maps to a public task instance.

\subsection{Prompts}
\label{app:prompt}

This section reproduces the stage prompts of the agentic rewriting backend
verbatim. They are the concrete instantiation of the repository transformation
in \LinkAlgorithm{alg:repo-transformation}: every stage is a separate
coding-agent session, and the session that implements a module never sees the
original source or the other modules. One rewriting cycle runs three stages:
\emph{decompose} (\LinkFigure{fig:prompt-decompose}), \emph{rewrite}
(\LinkFigure{fig:prompt-unit}), and \emph{compose}
(\LinkFigure{fig:prompt-compose}). A decomposition session reads the current
project and writes a unit manifest: a bottom-up decomposition into 3--8
core units, each with a responsibility, a behavioral specification, an
interface contract, the exact data strings its behavior fixes, and its
dependencies. Each unit is then implemented in isolation by a fresh session
that receives only that unit's entry and the interfaces of the units it
depends on. A composition session assembles the units into a working project
and iterates in a test loop against the behavioral conformance oracle, which
runs the source's captured golden cases on the evaluation machine, until
conformance is 1.0. The cycle then repeats on its own output, so the
decomposition is re-derived each cycle rather than inherited. After the final
cycle, a critic session (\LinkFigure{fig:prompt-critic}) removes the
source-chosen tokens that survived, gated on conformance staying at 1.0.
Braced placeholders are filled per unit or per cycle; every prompt is
otherwise sent to the agent exactly as printed. The headless-session driver
appends one further instruction to the compose and critic stages: ``Work
autonomously and keep iterating -- do not stop to ask.''

\paragraph{Decomposition prompt.}
The placeholder \texttt{\{target\}} is the source repository on the first
cycle and the composed repository on every later cycle, which is what makes
the decomposition re-derived rather than inherited.

\begin{center}
\begin{promptbox}
Decompose the Python project in {target}/ into its core modules so that each can be
rewritten independently by someone who cannot see the rest of the project.

Write units.json: a JSON array of the core units, ordered BOTTOM-UP so each unit depends only on
units that appear before it. Each entry:

{"name": "<short lowercase slug, used as the module name>",
 "layer": <int, 1 = depends on nothing internal>,
 "responsibility": "<one sentence>",
 "behavior": "<precise observable behavior: inputs, outputs, error categories, defaults, boundary
               and edge cases, data formats, ordering>",
 "interface": "<exactly what this unit exposes to the units above it: function and class names,
                signatures, and the meaning of every argument and return value>",
 "data_strings": ["<exact literal values this unit's behavior fixes>"],
 "depends_on": ["<names of units this one imports>"]}

Rules:
- Each unit must be fully implementable from its own entry alone. Its rewriter is told nothing about the rest of the project except the interfaces of the units it depends on.
- `interface` is the contract between units and the only channel of coordination between them.
- Together the units must cover every behavior the conformance cases exercise.
- Use 3-8 units. Fewer for a small project.
- Do NOT record file names, directory layout, or the original project's name anywhere.

Write only units.json.
\end{promptbox}
\captionof{figure}{Decomposition prompt. The session reads the current project and emits a
bottom-up unit decomposition with explicit interface contracts; no file names, directory layout,
or original project names may be recorded.}
\label{fig:prompt-decompose}
\end{center}

\paragraph{Isolated unit implementation prompt.}
One fresh session per unit, started in an empty directory containing nothing
but this prompt; the placeholder fields are filled from the unit's entry in
the unit manifest.

\begin{center}
\begin{promptbox}
Implement one module of a larger Python project, in this directory.

This is your entire specification. You cannot see the rest of the project.

UNIT NAME (this is the module name): {name}
RESPONSIBILITY: {responsibility}
BEHAVIOR: {behavior}
INTERFACE YOU MUST EXPOSE: {interface}
EXACT DATA STRINGS THAT MUST APPEAR IN OUTPUT: {data_strings}

Write this unit as a top-level Python module `{name}.py` (or a package directory `{name}/` with
`__init__.py`) in this directory, exposing exactly the interface above.

Dependencies: you may `import` these sibling modules, which other people are writing in parallel. Assume they exist at the project's top level and expose exactly these interfaces: {dep_interfaces}

Choose your own internal names, data structures and algorithms -- nothing here constrains them. Prefer the standard library; import a third-party package only if the behavior genuinely requires it. Write real, runnable code, no TODOs and no placeholder stubs.

Do not create files other than your unit's.
\end{promptbox}
\captionof{figure}{Isolated unit implementation prompt. The session sees only its unit's contract
and the interfaces of the units it depends on, never the source or the sibling modules.}
\label{fig:prompt-unit}
\end{center}

\paragraph{Composition prompt.}

\begin{center}
\begin{promptbox}
The units in units/ were rewritten independently, each by someone who could not see the others. Assemble them into a working Python project in repo/.

- Copy every unit's files in, preserving their module names.
- Resolve interface mismatches by adjusting the CALLERS to match what each unit actually exposes. Only rewrite a unit's internals if it plainly does not do what its contract says.
- Add the glue the project needs: the conformance entry point `repo/conf_cli.py`, and whatever wiring connects the layers.
- Build bottom-up: get the lowest layer importable and correct first, then each layer above it.

The entry point must read the JSON case array on stdin and write the JSON result array on stdout, and run as `python conf_cli.py`.

TEST LOOP: run `bash oracle.sh`. It copies repo/ to the evaluation machine and prints conformance plus the concrete cases that disagree. Iterate until conformance is 1.0 and n_failing is 0.
\end{promptbox}
\captionof{figure}{Composition prompt. The session assembles the independently rewritten units and iterates against the behavioral conformance oracle until every golden case passes.}
\label{fig:prompt-compose}
\end{center}

\paragraph{Critic prompt.}
Run once after the final cycle; \texttt{leaks.json} lists the source-chosen
tokens that survived the rewrite.

\begin{center}
\begin{promptbox}
repo/ still contains identifiers or strings that belong to ANOTHER project -- the one whose behavior this project reproduces. They are that project's own naming choices, not anything the behavior requires, and they must not appear in independently written code.

The surviving tokens are listed in leaks.json. Remove every one of them: rename identifiers to your own unrelated terms everywhere they occur, and reword any comment, message or documentation string containing them. Do NOT change observable behavior, and do NOT touch values the specification marks as externally required (protocol fields, third-party API keys, file-format field names). If one of these tokens appears in a value the conformance cases compare against, KEEP it -- behavior wins.

TEST LOOP: run `bash oracle.sh` after each substantial change. Conformance must stay 1.0.
\end{promptbox}
\captionof{figure}{Critic prompt. The literal list of surviving source-chosen tokens is fed back; a round is accepted only if the conformance oracle still passes every case.}
\label{fig:prompt-critic}
\end{center}

\section{Benchmark Characterization}
\label{app:characterization}

\subsection{Comparison with existing evaluations}
\label{app:benchmark-comparison}

Prior evaluations provide important fragment-level or transformation-level tests, but they do not combine a repository-scale, source-hidden agentic rewrite with a provenance task, an independent same-domain control, and abstention as an allowed decision. \LinkTable{tab:compare} summarizes these differences; the entries describe evaluation design rather than the general
capabilities of the underlying detectors.

\begin{table}[t]
\centering
\caption{\benchmark{} against related code-transformation and provenance evaluations. ``Source-hidden'' indicates that reconstruction proceeds without access to the source implementation; ``Independent control'' uses related but provenance-independent implementations; ``Abstention'' allows a detector to reject insufficient provenance evidence rather than force an attribution.}
\label{tab:compare}
\tablestyle\scriptsize
\setlength{\tabcolsep}{4pt}

\begin{tabularx}{\linewidth}{l*{6}{>{\centering\arraybackslash}X}}
\toprule
\rowcolor{TableHeader}
\tablehead{Benchmark} &
\tablehead{Repo. scale} &
\tablehead{Agentic rewrite} &
\tablehead{Source-hidden} &
\tablehead{Provenance task} &
\tablehead{Independent control} &
\tablehead{Abstention} \\
\midrule

BigCloneBench~\citep{svajlenko2014towards}
& -- & -- & -- & \checkmark & -- & -- \\

Mutation banks~\citep{duncan2021pyclone}
& -- & -- & -- & \checkmark & -- & -- \\

Adv.\ perturbation~\citep{yefet2020adversarial}
& -- & -- & -- & -- & -- & -- \\

SourceTracker/HST~\citep{gurioli2026sourcetracker}
& -- & -- & -- & \checkmark & -- & -- \\

RepoTransBench~\citep{wang2026repotransbench}
& \checkmark & \checkmark & -- & -- & -- & -- \\

RepoMod-Bench~\citep{li2026repomod}
& \checkmark & \checkmark & -- & -- & -- & -- \\

\rowcolor{TableFocus}
\benchmark{} (ours)
& \checkmark
& \checkmark
& \checkmark
& \checkmark
& \checkmark
& \checkmark \\

\bottomrule
\end{tabularx}
\end{table}

\paragraph{What each prior evaluation measures.}
\begin{itemize}
\item \textbf{BigCloneBench}~\citep{svajlenko2014towards} is a curated
benchmark of inter-project clone pairs mined from roughly $25{,}000$ Java
projects. Its unit is a single function, and pairs are tagged by clone type
from Type-1 through Type-4. The Type-3 and Type-4 labels rest on human
judgment over mined candidates, so the ground truth is least reliable exactly
for the semantic clones that a laundering benchmark must contain
\citep{krinke2022bigclonebench,krinke2025misuse}.
\item \textbf{Mutation banks}~\citep{duncan2021pyclone} generate clones by
applying a fixed catalog of syntactic mutation operators to a seed fragment.
The transformation space is bounded by the operator list, so a detector can
succeed by recognizing the operators rather than by finding provenance
evidence. Positives carry no behavioral check beyond the operator's
construction.
\item \textbf{Adversarial perturbation}~\citep{yefet2020adversarial} crafts
gradient-guided edits, such as identifier renamings, that flip a neural code
model's prediction on a snippet. This setting measures model robustness
rather than provenance: the transformation is minimal by design, the victim
model and ground truth are fixed in advance, and there is no notion of an
independent reimplementation against which a detector should abstain.
\item \textbf{SourceTracker/HST}~\citep{gurioli2026sourcetracker} retrieves
source snippets for verbatim or renamed fragments and is scored by Recall@N
and MRR. The unit is again a fragment, and these ranking metrics do not
test source-specific decisions against same-domain independent
repositories.
\item \textbf{Repository transformation
benchmarks}~\citep{wang2026repotransbench,li2026repomod} evaluate agentic
cross-language translation and repository modernization against executable
or implementation-independent behavioral tests. Both operate at repository
scale with agentic rewriting, but the agent reads the source during the
transformation, and the measured quantity is the functional validity of the
result rather than its provenance.
\end{itemize}

\paragraph{Why \benchmark{} differs.} Three requirements jointly separate
\benchmark{} from these settings. First, the unit is the repository and the
transformation is an agentic rewrite that proceeds without access to the
source implementation: the artifact is a source-hidden reimplementation
under a tested contract, not a mined pair, a catalog mutation, a gradient
perturbation, or a translation that reads the source. Second, the task is
provenance rather than functional validity, and every positive must pass an
execution-based conformance oracle on source-derived cases. Third,
calibrated detection and attribution use measured same-domain independent
references and permit abstention: ranking the true source first is not
sufficient for calibrated credit, and a detector may reject insufficient
provenance evidence. Native rows without validated bands remain explicitly
uncalibrated. The evaluations in \LinkTable{tab:compare} do not combine all
these requirements in a repository provenance task.

\subsection{Source repository statistics}
\label{app:source-stats}
\label{app:licenses}
\LinkTable{tab:licenses} summarizes archived GitHub license metadata for the
$1{,}510$ repositories represented in the benchmark: the $\nSources$
sources and $1{,}228$ further repositories from the negative list. The
metadata snapshots may postdate the pinned revisions, and an SPDX
identifier records a declared license, not a check of every file.

\begin{table}[ht]
\centering
\caption{Declared licenses from archived repository metadata. ``GPL family''
combines GPL, AGPL and LGPL identifiers; ``other SPDX'' combines MPL, EPL,
Unlicense, Creative Commons, 0BSD, WTFPL and MIT-0.}
\label{tab:licenses}
\tablestyle\footnotesize
\begin{tabularx}{\linewidth}{Lrr}
\toprule
\rowcolor{TableHeader}
\tablehead{License metadata} & \tablehead{Sources} & \tablehead{All required} \\
\midrule
Apache-2.0 & 120 & 389 \\
MIT & 50 & 410 \\
BSD-3-Clause / BSD-2-Clause & 21 & 51 \\
GPL family & 27 & 131 \\
Other SPDX & 5 & 25 \\
\midrule
SPDX identifier, total & 223 & 1{,}006 \\
NOASSERTION & 36 & 127 \\
No license field or metadata & 23 & 377 \\
\midrule
Repositories & \nSources & 1{,}510 \\
\bottomrule
\end{tabularx}
\end{table}

\begin{figure*}[t]
\centering
\includegraphics[width=\textwidth]{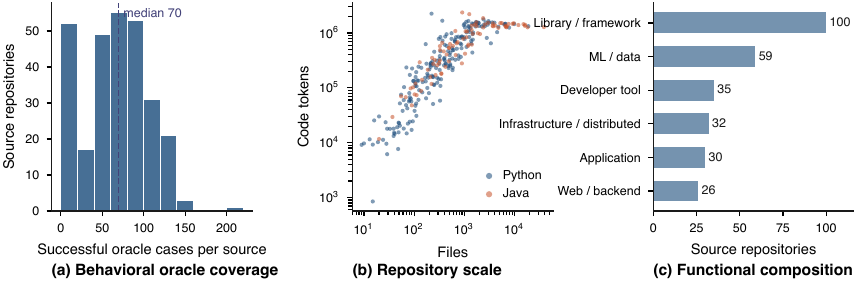}
\caption{Source repository evidence, scale, and composition over the
$\nSources$ released sources. (a) Number of successful behavioral oracle
cases available per source repository, summed over its verified artifacts
(median 70; 43 sources have no verified artifact). (b) Scale of the original
repositories: files versus code tokens (log--log), spanning 851 to
$2.4\times10^{6}$ tokens (median $3.5\times10^{5}$) and 9 to 38{,}532 files
(median 412). (c) Functional composition of source repositories across
libraries, applications, developer tools, ML/data systems, and
infrastructure.}
\label{fig:repo-stats}
\end{figure*}

\paragraph{Composition, scale, and behavioral evidence.}
\LinkFigure{fig:repo-stats} summarizes the source pool. Every source
repository is pinned to a specific commit in the released manifest, and all
construction and evaluation ran against those pinned revisions. The sources
are real, active projects and are large: originals span roughly three orders
of magnitude in size, with a median of $3.5\times10^{5}$ code tokens across
412 files. Repository scale is therefore itself a detection variable, which
\LinkSection{app:scale-scope} examines. Behavioral evidence backs the pool at
scale: verified artifacts carry 18{,}381 successful golden oracle cases in
total, a median of 70 per source, and every verified artifact clears the
$\pairOkGoldenMin$-case acceptance gate of
\LinkSection{app:verification}. Functionally, the pool is not concentrated in
any single domain: libraries and frameworks (100), ML and data systems (59),
developer tools (35), infrastructure and distributed systems (32),
applications (30), and web and backend projects (26), with the largest
category covering 35\% of sources.

\subsection{Functional preservation}
\label{app:functional-preservation}
\label{app:unseen-audit}
We audit 12 Pair instances from ten sources, with three instances from each
major language axis. Six new inputs were authored for each contract interface,
and none duplicates a visible conformance input. Original and rewritten code
ran in isolated copies. The inputs were authored by this project's working
agent, which is a limitation rather than a claim of independent testing.
\begin{table}[t]\centering
\caption{Bounded unseen-input audit on 12 Pair instances. Only inputs that
succeed on the original count toward agreement. Exact uses serialized JSON;
numeric equivalence additionally treats values such as 0 and 0.0 as equal.}
\label{tab:unseen-audit}\tablestyle\footnotesize
\begin{tabularx}{\linewidth}{Lrrrrr}\toprule
Axis & Sources & Inputs & Orig. success & Exact & Numeric equiv. \\\midrule
Python$\to$Python & 3 & 18 & 14 & 14 & 14 \\
Python$\to$JS & 3 & 18 & 15 & 13 & 13 \\
Java$\to$Java & 3 & 18 & 17 & 16 & 16 \\
Java$\to$JS & 3 & 18 & 17 & 17 & 17 \\
\bottomrule\end{tabularx}\end{table}
Of the 72 inputs, 63 succeed on the original and 60 of
these match exactly; numeric equivalence adds none. The three mismatches are
a dask Python$\rightarrow$JavaScript rewrite that rejects a valid
\texttt{evaluate\_graph} input and returns a dependency list in reversed
order, and a commons-jexl Java$\rightarrow$Java rewrite that tags an integer
result as \texttt{bigint} rather than \texttt{long}. This small deterministic sample tests
unseen inputs within generator-visible contract slices, not new functionality
or whole-project behavioral equivalence.

Static inspection of the 12 excluded entrypoints records their hashes and
named calls into included implementation modules. Nine handle stream/JSON I/O
and dispatch; AutoGPT, ComfyUI and commonmark-java also map operation or field
names. AutoGPT inserts argument defaults, and ComfyUI recursively substitutes
output strings and normalizes exceptions. Thus the adapters are not uniformly
pure byte transport: presentation and boundary behavior can be implemented in
excluded code. The audit identifies these responsibilities and delegation
boundaries; it does not certify all internal behavior or justify treating an
adapter token-ratio gate as a semantic proof.

\subsection{Rewrite magnitude and structural change}
\label{app:rewrite-magnitude}
\label{app:teaser-scores}
\LinkTable{tab:raw-score-summary} complements the matched comparison in
\LinkTable{tab:teaser-summary}. It retains each detector's native score scale:
statistical-profile scores can be negative, so they should not be placed
on the same radial scale as cosine or Jaccard similarities. For each
rewrite, we first take the median score across its measured same-domain
independent candidates; both displayed groups therefore contain one
observation per rewrite. Across-rewrite summaries can conceal within-query
ordering, which is why the teaser instead averages the fraction of
independent candidates outranked by the true source. Neither view treats
these same-domain controls as random repository pairs.

\begin{table}[ht]
\centering
\caption{Definition and interpretation of the native-score comparison over
all $\nPair$ benchmark rewrites. This preserves the evidential content of the
raw-score diagnostic without comparing incompatible detector scales.}
\label{tab:raw-score-summary}
\tablestyle
\begin{tabularx}{\linewidth}{lL}
\toprule
\tablehead{Element} & \tablehead{Definition} \\
\midrule
Rewrite--source group & One native detector score for every benchmark rewrite \\
Independent group & Within-rewrite median over measured same-domain candidates, then one observation per rewrite \\
Summary & Across-rewrite median and interquartile range; intervals are descriptive, not confidence intervals \\
Scale & Native per-detector scale; statistical-profile scores may be negative and are not pooled with cosine or Jaccard \\
Scope & Five repository-wide channels used in the matched diagnostic, not the complete clone-detector suite \\
\bottomrule
\end{tabularx}
\end{table}

\subsection{Information-access controls}
\label{app:independent-controls}
\label{app:t3-ladder}

\paragraph{Design.} This diagnostic compares the information supplied to the
rewriter on $29$ sources. Two contract-only implementations
(C$_{\mathrm{io}}$, two seeds) receive a purpose description, conformance
inputs, and golden outputs without source implementation access. The
source-visible B$_{\mathrm{inplace}}$ arm rewrites the implementation
in place; B$_{\mathrm{spec}}$ reuses a frozen specification-based rewrite.
The latter is a reference arm, not a new matched-generation run.
The contract-only arms receive source-derived outputs, so source hiding
alone does not establish provenance independence under
\LinkSection{app:task-construction}. We interpret them as information-access
controls, separately from the screened same-domain reference repositories.

\paragraph{Execution success is distinct from acceptance.}
All $58$ contract-only attempts execute, with median conformance $1.0$.
The source-visible arm executes on $15/29$ sources, with median conformance
$0.18$ among those runs. These execution counts are not acceptance counts.
The comparison concerns the tested recipes; it does not isolate source
visibility as the cause of behavioral failure.

\paragraph{Information access changes the residual similarity.}
The source-visible arm has a positive median margin advantage over the
contract-only arms on every channel (\LinkTable{tab:independent-controls}).
Separation is strongest for TF--IDF (AUROC $1.00$) and UniXcoder ($0.93$);
CodeBERT and GraphCodeBERT show weaker differences with $p>0.05$.
These comparisons use the $15$ executable source-visible outputs, rather
than $29$ behavior-verified positives. Band passage among the $58$
contract-only outputs is low for the lexical channels and reaches $13/58$
for GraphCodeBERT. Low band passage does not establish independent origin.

The frozen specification-regeneration arm has a negative median advantage
relative to the contract-only controls on every channel (UniXcoder
$-0.025$, $p=0.027$). The two contract-only implementations also resemble
each other (median TF--IDF similarity $0.28$) more than a source-visible
rewrite resembles its source ($0.17$). These observations support the
importance of supplied behavioral information, while the different
behavioral validity and generation histories limit causal interpretation.

\begin{table}[t]
\centering
\caption{Information-access controls on $29$ sources. $\Delta$ compares
source-visible and contract-only outputs on the 15 sources with executable
source-visible outputs, with Wilcoxon signed-rank $p$ and within-source AUROC.
Band pass counts the $58$ contract-only outputs above the source-specific
band. Executability does not imply acceptance; contract-only does not imply
provenance independence.}
\label{tab:independent-controls}
\tablestyle\footnotesize
\begin{tabularx}{\linewidth}{Lrrrr}
\toprule
\rowcolor{TableHeader}
\tablehead{Detector} & \tablehead{$\Delta$ median} & \tablehead{Wilcoxon $p$} & \tablehead{AUROC} & \tablehead{Band pass} \\
\midrule
TF--IDF & $+0.055$ & $6\mathrm{e}{-5}$ & $1.00$ & $0/58$ \\
k-gram & $+0.008$ & $1\mathrm{e}{-3}$ & $0.83$ & $5/58$ \\
Token Jaccard & $+0.002$ & $9\mathrm{e}{-4}$ & $0.87$ & $6/58$ \\
Stat. profile & $+0.549$ & $0.035$ & $0.67$ & $2/58$ \\
UniXcoder & $+0.088$ & $3\mathrm{e}{-4}$ & $0.93$ & $1/58$ \\
CodeBERT & $+0.001$ & $0.083$ & $0.55$ & $12/58$ \\
GraphCodeBERT & $+0.003$ & $0.135$ & $0.63$ & $13/58$ \\
\bottomrule
\end{tabularx}
\end{table}

\section{Detector Evaluation}
\label{app:detector-evaluation}

\subsection{Implementation details}
\label{app:impl-details}

\paragraph{Generator.}
Every release rewrite is produced by the Codex CLI 0.144.5 coding agent with
GPT-5.6-Sol as its default generator backbone, driven through the staged
contract, external-vocabulary and critic recipes of
\LinkAppendix{app:rewrite-protocol}. The generator study of
\LinkSection{app:generator-effects} re-executes the identical pipeline
with GPT-5.6-Sol and two alternative backbones, GLM-5.3 and
DeepSeek-V4.1-Flash, while holding the source set and evaluation protocol fixed.

\paragraph{How each detector family is tested.}
The lexical and statistical channels read the staged repository text
directly. The native clone tools inspect staged query and candidate repositories;
SourcererCC and MSCCD use recorded multi-document batches, while the
pairwise wrappers stage repositories side by side. Reported region pairs
are filtered to cross-repository matches and
aggregated into the token-coverage score of \LinkSection{app:pairlist}.
The embedding detectors encode a frozen windowed view of each repository
and pool the top query-window cosines
(\LinkAppendix{app:protocol-contract}). Every detector sees the same frozen
query and candidate repositories; unreadable inputs carry explicit
dispositions rather than zeros.

\paragraph{Java adaptations.}
StoneDetector, TACC and Oreo are Java-only, method-level tools, and we
adapted them rather than scoring them out of scope. Eligible Java files
are staged before any candidate comparison, so directory names cannot
shrink the coverage denominator. TACC assigns function IDs through an
unsynchronized static counter, so we run one native thread per comparison
and parallelize across isolated JVMs. MSCCD builds token frequencies
within each recorded batch, so its batch plans are retained as part of the
operating point. These three tools are reported only on their readable
Java$\to$Java inputs (\LinkAppendix{app:release-suite-detail}).

\subsection{Baselines and repository-level adaptation}
\label{app:baselines}
\label{app:protocol-contract}
\paragraph{Protocol boundary.}
The primary common-protocol results use the frozen score and band rules in
\LinkSection{sec:detectors}. Native clone records remain provisional.
This section distinguishes implemented input and aggregation rules from a
proposed source-held-out calibration contract. The proposed $1\%$ FPR
operating point and unique-top/unique-acceptance rules below do not define
the primary calibrated-attribution column.

\paragraph{Original method, adapter and task.}
We distinguish a method's original output from the repository adapter and
from the final decision rule. Clone detectors return accepted region pairs;
encoders return representations; a classifier may return only a label.
Region coverage is our adapter, not a probability or a score supplied by the
original detector. Encoder pooling is likewise a benchmark adaptation.
The comparison includes provisional repository adaptations of clone
detectors whose native outputs provide no repository ranking.
We report native detection and adapted attribution separately; neither
table is presented as a reproduction of the other task.
We record tool/checkpoint revisions, native thresholds, supported languages,
minimum unit sizes, preprocessing and resource limits. An AST-type-count
cosine surrogate cannot be reported as TreeCen's graph-centrality features
and trained classifier. UniXcoder requires its documented encoder-only
format; a GraphCodeBERT checkpoint without data-flow inputs is a
sequence-only ablation. Such substitutions must be labelled separately.

\paragraph{Inputs and applicability.}
Methods draw from the same frozen query and candidate repositories and
candidate sets, recorded with content hashes. We separately record each
preprocessor's effective file view, harness exclusions, parsing coverage
and native unit filters; shared repositories do not imply identical
inspected code. A Java-source-only encoder ablation isolates the effect of
including documentation and configuration files in the repository view.
The proposed source-held-out contract groups rewrites of one source and
known forks or duplicates, and requires disjoint tuning, calibration, and
evaluation source groups. Candidate-library reuse must also be audited.
These requirements do not certify that the primary source-conditional bands
implement a deployable source-held-out predictor. The existing random
control tier is not assumed to be an independent calibration set.
Matched sensitivity comparisons use the same valid queries within each
language axis. The primary native rows instead report their own readable
input populations, which need not coincide. Cross-language
stress tests are separate from a method's supported native scope.
Unsupported input, no eligible units, parse/mapping failure, timeout and
execution failure are distinct from valid execution with no match. They
are never converted to numerical zeros. Attribution requires 20 distinct,
valid candidate comparisons, with the source occurring exactly once.
We also report failures against the intended population fixed before
execution, with excluded IDs and reasons: the intersection of successful
runs alone can conceal failures on difficult inputs. Resource caps,
truncation, sampling and inspected-input fractions are recorded per method.
At this audit stage, all 1,382 referenced repositories agree with the
canonical host under the evaluated repository-text view. A separate native
raw-byte audit verifies all 2,526 input views after transport metadata removal,
including files beyond the text-view size cap. Neither audit establishes
absence of semantic duplicates.

\paragraph{Region-output adapter.}
The adapter proposed for validation is directional query coverage. Let
$U_d(q)$ be query token positions eligible under the declared input and unit
rules, fixed independently of the candidate, and let $M_d(q,c)\subseteq U_d(q)$
be their union over valid cross-repository matches. Define
\begin{equation}
 s_d(q,c)=\frac{|M_d(q,c)|}{|U_d(q)|}.
\end{equation}
An empty denominator is unavailable. Overlapping ranges count once; file
mapping and each tool's line-coordinate convention must be checked.
The denominator does not depend on candidate length, so adding unmatched
candidate code cannot dilute existing query evidence. We retain native
match/no-match outputs and report eligible-input fractions. Weighted
whole-repository coverage and symmetric coverage remain named sensitivity
analyses, not choices selected for favourable evaluation results.
In the implemented staged-Java sensitivity, file eligibility is fixed
before candidate comparison: all staged Java files up to 2,000,000 bytes.
The scorer does not apply another directory-name filter after native
staging. Thus a Java package named \texttt{env} contributes to the denominator
regardless of whether a candidate matches it. Multiline lexical tokens that
intersect several disjoint reported regions still count only once. These
checks repair aggregation; they do not validate the proposed native-unit
eligibility denominator.
For embedding adapters, model-specific input recipes precede a frozen
windowing and pooling rule. The diagnostic configuration uses 64
deterministically stride-selected windows, a 512-token limit and the mean
of the top eight query-window nearest-candidate cosines. All such settings
belong in representation-cache keys together with the checkpoint revision.

\paragraph{Execution integrity.}
Process success alone is insufficient: uncaught native worker failures can
leave a tokenizer waiting indefinitely or produce incomplete output. We
reject such output, retain errors and distinguish no eligible units from a
valid no-match result. For MSCCD, bounded batches retain every requested
repository pair; oversized pairs are not truncated. Native input/output maps,
configuration hashes and successful shard caches support replay. Runtime
changes to logging and queue polling require native-output equivalence
checks on positive and negative fixtures. Fixture agreement is reported as
a bounded regression check, not a universal equivalence claim.
MSCCD constructs token frequencies within each batch, so batch membership
is part of its operating point. We retain batch plans and do not infer
partition invariance merely from retaining every requested repository pair.
The unchanged TACC implementation assigns function IDs using an unsynchronized
static counter. A constructor-level concurrency probe reproduced duplicate IDs.
We therefore use one native thread per comparison and parallelize isolated JVMs.

\paragraph{Calibration without query labels.}
For each method and sufficiently populated language axis, the proposed
operating point targets an empirical false-positive rate of at most $1\%$
on calibration negatives. Require at least 100 negative pairs from 20
source groups; otherwise mark the axis uncalibrated. With the strict rule
$s>\tau$, choose the smallest observed-negative threshold satisfying this
constraint. This empirical constraint is not a population guarantee; report
realised counts and source-cluster uncertainty. Hyperparameter selection
uses tuning groups, while calibration sets only the threshold.
No evaluation label, true-source identity or test-optimal F1 threshold may
enter the prediction rule. The native/default operating point is reported
separately.

\paragraph{Attribution, ambiguity and abstention.}
To align scored repository adapters, the proposed primary decision returns
the uniquely highest-scoring candidate only if it exceeds its frozen
applicable threshold. Otherwise it abstains. We distinguish no accepted
evidence, tied highest scores and a highest-scoring candidate below its
threshold. We do not substitute a lower-scoring candidate because it passes
a different threshold. This requires validated comparability of candidate
scores within a query; different methods need not share a numerical scale.
Coverage scores are not provenance probabilities. This decision rule
remains subject to source-held-out validation.

As a separately named conservative sensitivity analysis, for a query's
fixed candidate set, form
\begin{equation}
 A_d(q)=\{c:s_d(q,c)>\tau_{d,\mathrm{axis}(q,c)}\}.
\end{equation}
Return the candidate only when $|A_d(q)|=1$. Otherwise abstain, distinguishing
no evidence ($|A_d(q)|=0$) from ambiguity ($|A_d(q)|>1$). This rule treats
candidates symmetrically and does not inspect the correct answer. All-zero
clone coverage gives abstention, not successful attribution. Report correct
and wrong attribution, abstention by reason, answer coverage and accuracy
conditional on answering; the last is unavailable at zero coverage.
Forced Rank@1 is a separate diagnostic for scored adapters, accompanied by
all-zero and top-tie rates. Fractional tie credit does not imply a successful
identification. A binary-only native output is not assigned an invented
continuous ranking. True-source rank times true-source band
passage remains labelled source-conditional verification.

This conservative unique-acceptance sensitivity rule is an explicit benchmark adapter,
not the original method's attribution algorithm. It abstains even if one of
several accepted candidates has a higher score. A unique-top or score-margin
rule requires a prespecified and calibrated adapter; the choice
cannot depend on challenge performance. Native hard-label methods retain
their accepted candidate sets and are reported separately from scored
coverage adaptations.

Pair-level FPR is not a query-level false-attribution guarantee. Even a
population per-pair bound $\alpha$ would yield only the union bound
$19\alpha$ on accepting any of 19 negative candidates; an empirical
calibration rate does not establish that population bound. We therefore
measure wrong attribution and ambiguity directly on held-out candidate
sets. Source-present tasks do not establish open-set rejection when the
true source is absent. The proposed operating point and minimum calibration
counts remain design choices under validation, not guarantees of
statistical power or query-level error control.

\subsubsection{Scoring clone-pair detectors at repository scale}
\label{app:pairlist}
This section documents the repository-level coverage adapter. The protocol in
\LinkAppendix{app:protocol-contract} separates native evidence,
eligible-query coverage, and calibrated prediction. Current SCC/MSCCD collection
uses recorded multi-document batches; it must not be described as independent
two-document runs.

The classical clone detectors of \LinkTable{tab:suite} do not return a
similarity between two repositories. Each returns a \emph{list of clone
pairs}: StoneDetector, TACC, Oreo and the AST surrogate emit
method-level pairs $(f_a, s_a, e_a, f_b, s_b, e_b)$ of file and line ranges;
MSCCD emits token-bag blocks with line ranges; SourcererCC in file-level mode
emits whole-file pairs, treated as the range of the whole file. The protocol
of \LinkSection{sec:detectors}, however, needs one number per (query,
candidate) pair for every metric it reports. The adaptation is a fixed
aggregator from a pair list to a score, applied identically to every tool.

\paragraph{AST surrogate, not native TreeCen.}
The AST-surrogate implementation counts named tree-sitter AST
node types per function and thresholds the cosine between count vectors at
$0.85$. It does not compute TreeCen's graph-centrality features or execute its
trained classifier. The archived upstream implementation instead constructs
tree graphs and centrality features before supervised clone classification.
Thus the AST-surrogate \texttt{treecen} scores, tables and neighbor sensitivity
measure this surrogate only. The local parser helper returns an empty vector
on exceptions and maps zero-norm comparisons to zero, without establishing
a valid no-match result. Native TreeCen evaluation requires a pinned feature
and classifier implementation, validated input/pair mappings and parsing,
and self-copy and partial-copy controls before its scores enter comparison.
The released audit materials record the
implementation hashes and a read-only check of the retained score tables.

\paragraph{Cross-repository matches only.} Pairwise wrappers stage a query
$A$ and candidate $B$ side by side. SourcererCC and MSCCD instead use the
recorded multi-document batch plans, retaining distinct project identifiers;
they are not independent two-document runs. Its output is
filtered to cross-repository pairs: a pair whose members both lie in $A$ or
both in $B$ is a within-repository clone (accessors, boilerplate, fixtures)
and says nothing about derivation. For native wrappers, filtering retains the reported cross-repository pairs; thresholds, minimum sizes and
verification stages are the recorded settings (StoneDetector LCS $0.3$,
minimum function size $10$; TACC $N{=}3$ with default seed and verify
scores; Oreo's released classifier; SourcererCC $70\%$; MSCCD $0.7$ with
$20$ minimum tokens; AST-surrogate cosine $0.85$).
The surrogate threshold is a local choice, not an upstream TreeCen default.

\paragraph{Token coverage.} Let $T(X)$ be the multiset of lexer tokens of
repository $X$ over the files the tool can read (each side lexed in its own
language), keeping keywords, names, literals, operators and punctuation and
dropping comments, whitespace and docstrings. For every file of $A$ the line
ranges of all surviving cross pairs are merged, and $M(A)$ counts the tokens
of $A$ inside a merged range; $M(B)$ likewise. The score is
\[
\mathrm{sim}(A,B) \;=\; \frac{M(A) + M(B)}{|T(A)| + |T(B)|},
\]
the fraction of the two repositories' code tokens that the detector places
inside some cross-repository clone. It is bounded in $[0,1]$ and symmetric;
the joint denominator prevents a small artifact from being rewarded for
covering its own few files against a large candidate; whole-file, block and
method pairs reduce to the same token scale, and a token is counted once
however many pairs cover it, so a tool that reports many small pairs gains
nothing. Only \emph{which} code a tool claims is cloned enters the score,
never a per-pair confidence: the object under test is the clone decision an
auditor would act on.

\paragraph{The self-pair ceiling.} On identical inputs a file-level tool
covers every token (SourcererCC $1.000$, MSCCD $0.9998$ on a Java source
against itself), whereas a method-level tool cannot: package and import
lines, field declarations, class headers and the methods it discards as too
small lie outside every pair, so the same self-pair scores $0.48$
(StoneDetector), $0.38$ (TACC) or $0.73$ (Oreo). This ceiling depends on the tool and its adapter, so raw coverage scores
are not directly comparable across tools. Source-specific bands are used
only where validated for the reported protocol; the clean-input native rows
in \LinkTable{tab:suite} have no calibrated results. AUROC and best F1 remain
within-detector summaries over readable inputs.

\paragraph{Alternatives not used.} An earlier harness scored
\emph{function coverage}, $\tfrac12(\lvert\text{covered units of }A\rvert/\lvert\text{units of }A\rvert + \lvert\text{covered units of }B\rvert/\lvert\text{units of }B\rvert)$
over methods extracted by the benchmark's own parser. It weights a three-line
accessor and a three-hundred-line parser equally, depends on an extractor
external to the tool, and is undefined for whole-file tools; token coverage
replaced it in the pairlist harness. Aggregation does not validate the tool
or the mapping from reported pairs to source ranges. No pair
is weighted by the tool's similarity value and no per-tool threshold is tuned
on the benchmark. Every document is staged without the conformance harness
and build outputs. Effective file views and eligible units still differ
across native tools, construction channels, and embedding adapters.

\subsection{Complete Pair results}
\label{app:pair-results}
\label{app:native}
The benchmark's three construction channels (lexical \texttt{tfidf\_cosine},
character-level \texttt{kgram\_jaccard}, statistical \texttt{stat\_profile}) and the
token-overlap measure \texttt{token\_jaccard} are deliberately simple,
unpublished instruments: they are the channels the construction itself uses
for difficulty scoring and band calibration, and they double as strong
sanity baselines. This section reports their full results.

\subsubsection{Complete detector record}
\label{app:release-suite-detail}
This section collects every detector record on the
release in two blocks. The upper block holds the eight band-calibrated
records on all Pair and Attr instances: the three encoders of
\LinkTable{tab:suite}, the three construction channels, token Jaccard and an
exploratory AST surrogate (the \texttt{treecen} record, labelled
explicitly throughout the statistical appendices). The lower block holds the
five native clone detectors as provisional records on their readable
inputs: SourcererCC and MSCCD read 368 and 363 Pair
positives and 69 and 64 Attr queries; the Java-only
StoneDetector, TACC and Oreo read 60, 63 and 63
Java$\rightarrow$Java positives and 16, 16 and 11 of that axis's Attr
queries. Every unread instance carries an explicit disposition. Neither
block is a validation of published detectors, and the provisional values
are not comparable with the band-calibrated records or with one another.

Among the eight band-calibrated records, calibrated TPR on the $\nPair$ Pair positives ranges from
$0.009$ to $0.124$, while pooled AUROC ranges from $0.288$ to $0.661$.
Thus the benchmark degrades ranking as well as calibrated decisions;
it does not support a claim that all ranking metrics remain near ceiling.
The three highest band-pass rates require their negative-pair context: the
AST surrogate exceeds the band on $12.4\%$ of positives but
also on $12.8\%$ of same-domain negative pairs, the sequence-only
GraphCodeBERT on $10.4\%$ against $5.8\%$, and CodeBERT on $7.4\%$
against $3.4\%$. The AST surrogate thus has nearly equal pooled TPR and FPR;
the two encoders have a positive TPR--FPR gap, although all three yield
conjunctive attribution on only $1$--$2\%$ of Attr instances. Among the
band-calibrated records, UniXcoder retains the strongest attribution ranking
signal (Rank@1 $0.220$), but conjunctive calibrated attribution is only
$0.010$.

The distinction between ranking and decisions remains informative even
when neither is strong. TF--IDF ranks the source first on $5\%$ of Attr
instances, but none of those instances also clears its band. Token
Jaccard crosses the source band on $6\%$ of attribution queries, yet never ranks the
source first; counting band passage as attribution would therefore give a
misleading success rate. We recompute the final attribution column from
the recorded top-rank/tie information and band decisions, requiring both.
The small nonzero k-gram value is fractional credit from one above-band
13-way top tie, not an unambiguous source identification.

The axis-level records show that TF--IDF band passage is
$19/78$ on Java$\rightarrow$Java and zero on the other axes, and the same
axis supplies $7$ of UniXcoder's $10$, $29$ of CodeBERT's $32$ and $39$ of
GraphCodeBERT's $45$ accepted positives. These are not only pairs the
construction channels already flag: $2$ of UniXcoder's $7$, $19$ of
CodeBERT's $29$ and $24$ of GraphCodeBERT's $39$ Java$\rightarrow$Java
acceptances fall among the $55$ pairs of that axis below the band on every
construction channel. The encoders thus add evidence where the channels see
none, but on this axis CodeBERT and GraphCodeBERT also accept $13.2\%$ and
$24.3\%$ of same-domain negative pairs. No single rewriting recipe drives
the result: over the full-contract, external-vocabulary and critic recipes
($148$, $139$ and $146$ positives), accepted positives are $15/18/12$ for
GraphCodeBERT, $13/12/7$ for CodeBERT, $2/5/3$ for UniXcoder and $8/7/4$ for
TF--IDF, differing by at most six for every detector without a consistent
ordering.
\LinkAppendix{app:release-statistics} gives the upper-block records other
than the sequence-only GraphCodeBERT by major axis
(\LinkTable{tab:axis-0}).

The five provisional native records were collected under the clean-input
protocol without reference bands: their pooled AUROC spans $0.501$
(SourcererCC) to $0.632$ (StoneDetector), and forced Rank@1 spans $0.000$ to
$0.312$ (TACC) over their readable attribution queries. These are the native entries of \LinkTable{tab:suite},
with coverage not comparable with its common-protocol rows: Oreo's AUROC
rests on $63$ Java$\rightarrow$Java positives, the encoders' on $\nPair$
positives across four axes. Their band-calibrated columns are dashes rather
than zeros because that clean-input protocol has no validated bands; the native-output
mapping audits and the parsed-units sensitivity
(\LinkAppendix{app:protocol-contract}) govern their validity boundary.
Deleting the most influential calibration neighbor raises both the AST surrogate's TPR
and FPR on eligible queries (\LinkAppendix{app:neighbor-sensitivity});
the empirical band is sensitive to neighborhood composition.

\begin{table}[t]
\centering
\caption{Relative evidence versus calibrated decisions on all $\nPair$
Pair instances. True-source superiority averages, per rewrite, the fraction of
measured same-domain candidates scored below the true source; ties receive half
credit. Intervals use $4{,}000$ source-cluster bootstrap resamples.}
\label{tab:teaser-summary}
\tablestyle
\begin{tabularx}{\linewidth}{lLL}
\toprule
\tablehead{Result} & \tablehead{Observation} & \tablehead{Consequence} \\
\midrule
Relative ordering & TF--IDF and UniXcoder retain source preference; three other point estimates fall below the 0.5 no-preference reference & Some relative signal survives rewriting \\
Calibrated detection & Only 0.9--4.4\% clear the detector-specific independent-code band & Relative preference rarely becomes actionable evidence \\
Rewrite validity & Names, modules, and sometimes language change while oracle-covered behavior remains & A new implementation is not evidence of independent origin \\
\bottomrule
\end{tabularx}
\end{table}

\LinkTable{tab:teaser-summary} makes the detection problem concrete. Across the
$\nPair$ benchmark rewrites, k-gram, statistical-profile and token-Jaccard
scores favor the true source over same-domain independent candidates less
than half the time, with ties receiving half credit. TF--IDF and UniXcoder
retain relative signal, but only $0.9$--$4.4\%$ of rewrites exceed the
source-specific independent-code band on any one of these five channels.
The finding is therefore not that every similarity signal disappears:
relative similarity can survive without exceeding the empirical source-group
threshold. This is a failure under that decision rule, not an absence of all
possible provenance evidence. These values characterize the benchmark and
are not a success rate for arbitrary rewriting attempts. \LinkSection{sec:detectors}
reports the encoder comparison and the distinction between ranking
and grouped-band decisions.

\subsubsection{Stratified and axis-level results}
\label{app:stratified-results}
\label{app:release-statistics}
All statistics in this section reuse the frozen per-instance records; no
model, detector or behavioral test was rerun. Pair has $\nPair$ instances
from $\nSources$ sources and Attr has $\nAttr$ from $\nAttr$. Pair quality
tiers span hard, borderline, medium and stub: the single stub-tier instance
(\texttt{java-jwt}, capped size ratio 0.187) enters
through the Java$\rightarrow$Java axis and has
$d_{\mathrm{pair}}>0$, so it is not in the challenge tier. For each task, we draw
4,000 bootstrap samples of its sources with replacement, retaining all
instances of each sampled source and recomputing the instance-weighted rate
(NumPy PCG64, seed 20260916, sorted source names). Brackets give percentile
95\% intervals conditional on the fixed artifacts, bands, and task manifests.
They do not account for generation, neighbor selection, or unseen-source bias;
zero-success intervals are degenerate bootstrap intervals, not population
upper bounds. We use stored decisions and exact top-tie counts rather than
rethresholding rounded scores. AUROC, best-F1, FPR and Rank@1 intervals are
not estimated here. The AST surrogate is the \texttt{treecen}
record, not the published TreeCen method; score completeness does not certify
native-tool validity.
\begin{table}[p]\centering
\caption{Detector performance across the four major rewriting axes.
Same-language Java retains substantially more relative provenance signal
than the other axes, but calibrated detection remains limited and often
comes with increased false-positive rates. Cross-language rewriting
amplifies provenance loss but is not necessary for it. Calibrated attribution
(omitted from the columns) is zero except for CodeBERT on Java-to-Java
($0.050$) and k-gram on Java-to-JavaScript ($0.004$). TPR and calibrated
attribution use exact stored decisions and tie counts; AUROC/FPR are the
frozen per-axis summaries.}
\label{tab:axis-0}\tablestyle\footnotesize
\begin{tabularx}{\linewidth}{Lrrrr}\toprule
Detector & AUROC & TPR & FPR & Rank@1 \\\midrule
\tablegroup{5}{Python to Python}
UniXcoder & 0.635 & 0.000 & 0.000 & 0.250 \\
CodeBERT & 0.574 & 0.000 & 0.000 & 0.050 \\
AST surrogate & 0.487 & 0.150 & 0.277 & 0.000 \\
TF--IDF & 0.581 & 0.000 & 0.000 & 0.000 \\
k-gram & 0.415 & 0.000 & 0.000 & 0.000 \\
Stat. profile & 0.477 & 0.000 & 0.038 & 0.000 \\
Token Jaccard & 0.341 & 0.000 & 0.057 & 0.000 \\
\tablegroup{5}{Python to JavaScript}
UniXcoder & 0.625 & 0.000 & 0.000 & 0.250 \\
CodeBERT & 0.532 & 0.000 & 0.000 & 0.050 \\
AST surrogate & 0.586 & 0.000 & 0.013 & 0.050 \\
TF--IDF & 0.603 & 0.000 & 0.000 & 0.050 \\
k-gram & 0.447 & 0.000 & 0.000 & 0.000 \\
Stat. profile & 0.444 & 0.000 & 0.000 & 0.000 \\
Token Jaccard & 0.339 & 0.000 & 0.025 & 0.000 \\
\tablegroup{5}{Java to Java}
UniXcoder & 0.834 & 0.090 & 0.000 & 0.150 \\
CodeBERT & 0.705 & 0.372 & 0.132 & 0.050 \\
AST surrogate & 0.645 & 0.385 & 0.207 & 0.050 \\
TF--IDF & 0.708 & 0.244 & 0.033 & 0.050 \\
k-gram & 0.405 & 0.051 & 0.126 & 0.025 \\
Stat. profile & 0.539 & 0.077 & 0.081 & 0.000 \\
Token Jaccard & 0.251 & 0.179 & 0.237 & 0.000 \\
\tablegroup{5}{Java to JavaScript}
UniXcoder & 0.611 & 0.000 & 0.014 & 0.100 \\
CodeBERT & 0.558 & 0.030 & 0.002 & 0.000 \\
AST surrogate & 0.503 & 0.000 & 0.022 & 0.050 \\
TF--IDF & 0.538 & 0.000 & 0.008 & 0.000 \\
k-gram & 0.389 & 0.000 & 0.004 & 0.011 \\
Stat. profile & 0.429 & 0.000 & 0.000 & 0.000 \\
Token Jaccard & 0.219 & 0.000 & 0.081 & 0.000 \\
\bottomrule\end{tabularx}\end{table}
\paragraph{Strict-positive sensitivity.} Restricting Pair to full positive-case conformance leaves $62\%$ of instances. Of these, $45\%$ meet the challenge rule ($d_{\rm pair}<0$ and capped size ratio $\geq0.2$). These are checks on the same generation-visible cases, not unseen tests.

\subsection{Complete Attribution results}
\label{app:attr-results}
\label{app:attr-audit}
The main-table values were recomputed from the frozen per-query records,
using the exact counts of candidates above and tied with the true source.
All $\nAttr$ benchmark queries have $20$ distinct candidates and contain the
true source exactly once. A Rank@1 of $0.050$ can arise from a small number
of unique wins or from uninformative $20$-way ties; these are distinct
outcomes. \LinkTable{tab:attr-diagnostics} separates them.

\begin{table}[t]\centering
\caption{Attribution diagnostics. Unique wins, top ties, all-candidate ties
and band passes are fractions of the $\nAttr$ attribution queries; band
passage alone is not successful attribution. Rank@1 and source-band
attribution are means of fractional credits.}
\label{tab:attr-diagnostics}\tablestyle\footnotesize
\begin{tabularx}{\linewidth}{Lrrrrrr}\toprule
Method & Unique wins & Top ties & All-20 ties & Band passes & Rank@1 & Band credit \\\midrule
UniXcoder & 0.22 & 0.00 & 0.00 & 0.03 & 0.220 & 0.010 \\
CodeBERT & 0.06 & 0.00 & 0.00 & 0.11 & 0.060 & 0.010 \\
AST-type-count surrogate & 0.05 & 0.00 & 0.00 & 0.14 & 0.050 & 0.010 \\\midrule
TF--IDF & 0.05 & 0.00 & 0.00 & 0.04 & 0.050 & 0.000 \\
k-gram & 0.00 & 0.03 & 0.00 & 0.02 & 0.007 & 0.001 \\
Statistical profile & 0.00 & 0.00 & 0.00 & 0.02 & 0.000 & 0.000 \\
Token Jaccard & 0.00 & 0.00 & 0.00 & 0.06 & 0.000 & 0.000 \\\bottomrule
\end{tabularx}
\end{table}

The earlier 14-query SourcererCC and MSCCD pilot is different: every query
has zero similarity to all 20 candidates. Their $0.050$ forced Rank@1 is
solely the expected credit under random tie-breaking; neither has a
successful calibrated attribution. It is not evidence of useful source
identification. Those pilot records also predate the stricter subprocess
checks and are not used to validate the current tool installation.

\paragraph{Adaptation and decision scope.}
For clone-pair methods, attribution is a benchmark adaptation: detected
regions become repository-level token-coverage scores, which rank the 20
candidates. This is not an attribution algorithm supplied by the original
clone detector. Valid execution and readable inputs must be established
before interpreting zero coverage as a detector outcome; unsupported
languages and failed executions are not negative evidence.
The reported calibrated credit tests the true source's rank and reference
band. The stored records do not contain calibrated decisions for every
incorrect candidate, so they do not determine an operational abstention
rate or false-attribution rate. These require candidate-specific calibration
and a fully specified sampling-and-abstention rule.

\paragraph{Result-field semantics.}
In the frozen evaluator records, the stored attribution flag denotes band
passage, not conjunctive attribution. The paper's calibrated-attribution
column is instead computed as top-rank credit times that flag. The
released records expose band passage and calibrated attribution
separately, and retain all 20 unrounded candidate scores for audit. A rounded similarity
must never be used to reconstruct a near-threshold decision.

The AST-surrogate record labelled treecen is an AST-type-count cosine
surrogate, not a faithful implementation of TreeCen. Its full 20-candidate score records
reproduce the stored rank arithmetic, but $18\%$ of queries contain at least
one candidate with empty or otherwise invalid token evidence.
Among queries without those invalid comparisons, weighted and query
coverage yield different
top-candidate sets in $72$ of $82$ cases. Both have unique winners on $30$
queries; $20$ of these winners differ. This sensitivity
audit does not select an aggregation or validate the original detector.

\paragraph{Transport metadata audit.}
Repositories are transferred between machines, and transfer can introduce
macOS AppleDouble sidecar files; 561 transferred rewrite directories
contained them. The input views exclude a sidecar only when both its
dot-underscore filename prefix and its AppleDouble magic bytes agree;
genuine hidden source files are retained, and the broad repository-text
loader rejects these non-UTF-8 files. Transport metadata is not benchmark
source code. All 2,526 native input views are byte-verified against the
canonical sources, covering 1,382 reference repositories and 1,144
distinct rewrite paths, and all native records reported in this paper use
these verified views.

\paragraph{Encoder input-view sensitivity.}
The Java-source ablation uses the verified inputs. On 46 matched
queries from 19 source IDs, CodeBERT identifies the unique top source in
4 cases with Java-source input and 6 with the repository-text view;
UniXcoder does so in 19 and 14 cases, respectively
(\LinkTable{tab:java-input-ablation}). Twelve of the 58 requested queries
lack a complete 20-candidate Java-source view and are excluded explicitly.
The direction differs across encoders and the intervals overlap, so no view
is uniformly better. These are unthresholded, source-present ranking diagnostics, with no
abstentions on the matched cohort; they do not establish calibrated
attribution or determine which input view is primary. Intervals resample
source IDs, whose fork independence has not been certified.
\begin{table}[t]\centering
\caption{Input-view sensitivity for the encoder adaptations on 46 matched
Java attribution queries from 19 source IDs. All queries retain the same 20
candidates. Accuracy counts unique correct top scores without a threshold;
intervals are descriptive 95\% source-ID bootstrap percentile intervals.}
\label{tab:java-input-ablation}\tablestyle\footnotesize
\begin{tabularx}{\linewidth}{Lrrr}\toprule
Adaptation and input & Correct & Accuracy & Interval \\\midrule
CodeBERT, repository view & 6/46 & 0.130 & [0.021, 0.280] \\
CodeBERT, Java source & 4/46 & 0.087 & [0.000, 0.200] \\
UniXcoder, repository view & 14/46 & 0.304 & [0.143, 0.478] \\
UniXcoder, Java source & 19/46 & 0.413 & [0.212, 0.622] \\
\bottomrule\end{tabularx}
\end{table}

\subsubsection{Paired source-absence diagnostic}
\label{app:source-absence-diagnostic}
\LinkTable{tab:source-absence-diagnostic} reports computed decisions under
frozen per-axis thresholds for 20 benchmark queries. For each
source-present query we remove its true source, retain all 19
distractors, and apply the same threshold and unique-top prediction rule.
Labels enter construction of this sensitivity condition and outcome
assessment, never the prediction function. This is a 19-candidate paired
removal experiment, not a new 20-candidate open-set benchmark or an
independent population of unrelated queries. Thresholds are not refitted.
The diagnostic covers 15 Java-to-Java and five Python-to-Python queries. It
separates erroneous answers from abstention, but does not yet cover all
$\nAttr$ attribution queries.
\begin{table}[t]\centering
\caption{Source-present and paired source-absent decisions on 20 benchmark
attribution queries. The decision rule is unique-top attribution with frozen per-axis thresholds;
present-source counts use 20 candidates; the paired absent-source
sensitivity removes the true source and retains 19 distractors without
recalibration. Correct/wrong/abstained (C/W/A) and false attribution/
correct rejection (FA/CR) are query counts.}
\label{tab:source-absence-diagnostic}\tablestyle\footnotesize
\begin{tabularx}{\linewidth}{LLrrrrrr}\toprule
Method & Axis & $n$ & C & W & A & FA & CR \\\midrule
CodeBERT & Java & 15 & 0 & 5 & 10 & 5 & 10 \\
CodeBERT & Python & 5 & 0 & 0 & 5 & 0 & 5 \\
UniXcoder & Java & 15 & 0 & 1 & 14 & 1 & 14 \\
UniXcoder & Python & 5 & 1 & 0 & 4 & 1 & 4 \\
SourcererCC & Java & 2 & 0 & 0 & 2 & 0 & 2 \\
SourcererCC & Python & 1 & 0 & 0 & 1 & 0 & 1 \\
\bottomrule\end{tabularx}
\end{table}

\subsection{Language-specific / Java-tier results}
\label{app:java-tier}

\subsubsection{Provisional Java-axis record: StoneDetector}
\label{app:stone-audit}
An extended StoneDetector pairlist, separate from the clean-input row in
\LinkTable{tab:suite}, covers the evaluated Java-to-Java Pair
and Attr IDs in full, making it applicable-axis complete. Under the
zero-on-error/unsupported missing-score policy it records AUROC $0.716$ and
best F1 $0.515$, with Pair TPR $16/78$ and FPR $5/334$, Rank@1 $1/20$ and
calibrated attribution $0/20$ under the conjunctive rule, alongside the
same-axis comparison in \LinkTable{tab:axis-0}. This is an audit of
recorded outputs, not certification
of a validated baseline: unsupported comparisons and timeouts are stored as zero,
so ranking and band construction need explicit missing-score handling.
All ten errors are timeouts. Four are source--neighbor comparisons for
\texttt{elasticsearch}, outside these evaluated cohorts. One is the
\texttt{mvel}--\texttt{lombok} calibration comparison; three are the
\texttt{mvel} rewrite--\texttt{lombok} negative pairs. The remaining two
are Attr distractor comparisons (\texttt{commonmark-java} against
\texttt{opensearch}, and \texttt{commons-jxpath} against \texttt{hadoop}).
No evaluated positive comparison timed out. Removing queries with any timed-out
positive, negative, calibration or candidate comparison leaves Pair TPR
$16/75$ and Attr $0/18$. This exclusion is a sensitivity check, not an estimate
of performance on the omitted queries.

Unsupported records affect at least one required comparison for 15 Pair
queries and four Attr queries. Their zeros are not evidence of a valid
no-match result. The raw Pair false-positive count is $5/334$ under this policy. Neither this provisional record nor the timeout exclusion
establishes correct parsing, local-copy recall, or robustness to the whole-repository
token denominator. Applicable-input self-copy and partial-copy checks,
explicit unsupported/error outcomes, and rerunning affected comparisons are
needed before treating this record as a validated baseline.

\section{Explanatory Analyses}
\label{app:analyses}

\subsection{What provenance signals survive rewriting?}
\label{app:surviving-signals}

\subsubsection{What the surviving lexical similarity is made of}
\label{app:s2-contract}
\paragraph{Design.} This diagnostic splits token-frequency cosine between
a source and artifact into contributions from contract vocabulary and other
tokens. It uses term-frequency weighting without IDF, so the additive split
is exact for this diagnostic, not for the primary TF--IDF detector.
A token is assigned to the contract component when it occurs in the
frozen contract material of that instance (the purpose statement, the
benchmark test cases, or the golden outputs) and to the incidental
component otherwise; the split is computed per instance and the two
components sum to the full cosine with residual
$\le 5.6\times10^{-17}$ over the evaluated instances on the benchmark's four
rewriting axes.
The token assignment uses no learned model or fitted parameter.

\paragraph{Results.} The separately summarized component medians are
nearly equal ($0.143$ contract and $0.143$ incidental), while the median
per-instance contract share is $0.530$. The contract component is larger
on $55.6\%$ of instances. The share is not uniform across the
benchmark's axes: it is highest where source and artifact are written in
different languages (py$\rightarrow$js
$0.67$, java$\rightarrow$js $0.57$) and lowest on the same-language
axes (py$\rightarrow$py $0.40$, java$\rightarrow$java $0.48$), the ordering
expected if cross-language rewrites retain contract vocabulary while
discarding incidental identifiers along with the syntax
(\LinkTable{tab:app-contract-share}).

The split also provides a diagnostic of the encoder margin. The
contract share is \emph{negatively} associated with the UniXcoder
margin over the same instances (Spearman $\rho=-0.227$,
$95\%$ CI $[-0.327,-0.117]$;
\LinkFigure{fig:rewrite-depth}\hyperref[fig:rewrite-depth]{b} plots the
two quantities per instance over the four axes): higher contract share is associated with a lower UniXcoder margin.
The diagnostic's full cosine correlates with the frozen TF--IDF score at
$\rho=0.353$, further distinguishing their numerical scales. These
associations are consistent with functional vocabulary contributing to
residual similarity, but do not identify which features the encoder uses
or establish a causal mechanism.

\begin{table}[t]
\centering
\caption{Contract share of token-frequency cosine, by axis. The share is the
fraction of the normalized dot product contributed by tokens occurring
in the instance's frozen contract material; the two components sum to
the full cosine exactly.}
\label{tab:app-contract-share}
\tablestyle
\begin{tabularx}{\linewidth}{Lccc}
\toprule
\rowcolor{TableHeader}
\tablehead{Axis} & \tablehead{Median contract share} & \tablehead{IQR} & \tablehead{Median diagnostic cosine} \\
\midrule
py$\rightarrow$js     & 0.67 & $[0.56, 0.77]$ & 0.17 \\
java$\rightarrow$js   & 0.57 & $[0.51, 0.63]$ & 0.27 \\
java$\rightarrow$java & 0.48 & $[0.38, 0.56]$ & 0.44 \\
py$\rightarrow$py     & 0.40 & $[0.26, 0.47]$ & 0.42 \\
\bottomrule
\end{tabularx}
\end{table}

\subsubsection{Deterministic channels predict failure without providing detection}
\label{app:t8-channels}
\paragraph{Design.} The study was pre-registered and the analysis plan
sealed before computation. The three deterministic diagnostic
channels (identifier, string-literal, import-call overlap, extracted
with tree-sitter) are summarized per instance by their channel margins
(score minus per-source band). A logistic probe on the three margins is
trained to predict which benchmark artifacts pass a detector's band,
with folds fixed to the released leave-one-source-group-out audit splits;
no hyperparameter is tuned on any test fold. The pre-registered folds
cover $37$ source groups ($154$ instances); the
remaining sources lie outside every fold and are excluded from both
training and test. Decisions use the same grouped-band rule as the
benchmark: the pair $(A,B)$ is called detected when the probe score
exceeds the maximum probe score over $A$'s same-domain negatives.
Leave-one-channel-out refits report which channel carries the
prediction.

\paragraph{Results.} The margins alone predict band passage. For the
lexical (tfidf) target, which has $12/154$ positives, the probe reaches
held-out AUROC $0.80$; dropping the identifier channel costs the most
($0.70$), dropping string literals $0.74$, dropping import calls
nothing ($0.80$). For the encoder (UniXcoder) target, with $4/154$
positives, AUROC is $0.75$ and flat under every leave-one-channel-out
refit, consistent with the encoder signal not being concentrated in any
one deterministic channel. The grouped-band decisions tell the other
half: the probe recovers only $15/154$ (lexical) and $13/154$
(encoder) instances, barely above the detectors' own $12$ and $4$. Predicting a detector's band outcome is different from establishing
provenance. The fitted probe recovers some additional positives, but these
counts alone do not establish source-specific detection without its
false-positive context. In particular, without the identifier
channel the grouped-band recovery \emph{rises} to $31/154$: identifier margins are also high for same-domain negatives. This is
consistent with shared vocabulary helping predict detector behavior without
necessarily improving a calibrated provenance decision.

\subsection{Structural and rung-wise rewrite evidence}
\label{app:rewrite-evidence}

\paragraph{Structural retention.}
The frozen descriptor study records source-to-artifact structural similarity
for $1{,}156$ source-artifact pairs spanning the rewriting rungs. Median
retention is $0.156$ for file count, $0.011$ for repository size, $0.141$ for
AST-trigram cosine (400 pairs with complete parses), $0.008$ for identifier
Jaccard, $0.005$ for string-literal Jaccard, and $0.003$ for import/call
similarity
(\LinkFigure{fig:main-agentic}\hyperref[fig:main-agentic]{a}). These descriptors record broad implementation changes, but aggregate
similarities do not establish that every artifact changes every channel.
A file-count ratio measures the number of files, not the fraction of original
files retained; import/call similarity is not a dependency-DAG identity test.

\paragraph{Rung-wise decay.}
Restricting to fully conforming artifacts ($n=101$, $85$, and $79$ across the
full-rewrite, contract, and critic rungs), the embedding detectors are
already usually below the source band at the mildest rung. The median UniXcoder
margin is $-0.19$, and its band-pass rate is $1.0$--$3.5\%$ across rungs.
Only the lexical TF--IDF channel retains a gradient. Its band-pass rate falls
from $0.376$ to $0.341$ to $0.203$ across the three rungs
(\LinkFigure{fig:main-agentic}\hyperref[fig:main-agentic]{b}). The critic rung feeds surviving source names back to the agent.
Its accepted artifacts have lower lexical band passage, but the rung
populations differ, so this comparison does not isolate the critic's effect.

\paragraph{Generator validity and evasion.}
Under the identical pipeline, the valid rewrite rate rises from $0.875$
(GLM-5.3) through $0.925$ (DeepSeek-V4.1-Flash) to $0.950$ (GPT-5.6-Sol),
while evasion among valid artifacts stays between $0.83$ and $0.95$
(\LinkFigure{fig:main-agentic}\hyperref[fig:main-agentic]{c}).
\LinkSection{app:generator-effects} reports the per-model records.

\subsection{Rewrite strength and detection difficulty}
\label{app:strength-difficulty}

\paragraph{Design.} One round of the agentic pipeline is already a full
behavior-constrained reconstruction: the repository is decomposed into a
specification and units, each unit is rewritten in isolation, and the
units are composed against the source's conformance oracle. Chaining
three rounds, where every round re-decomposes the repository produced by
the previous round while the oracle stays fixed to the original source,
gives a controlled ladder of rewrite depth: behavior is pinned to the
same tests while the artifact is pushed further away from the source's
surface. We run three rounds on the $40$ Python sources of the agentic
pool with the same rewriting model as the main agentic runs, score every
accepted round (conformance $\ge 0.99$) with the frozen detector suite
and the deterministic channel margins, and ask a single question: does
provenance evidence collapse at the first reconstruction, or does it
keep decaying while behavior is preserved? The analysis is fixed in
advance: median margins per depth, and paired Wilcoxon signed-rank tests
between consecutive depths over chains complete at both, with
saturation claimed only when the drop from depth $1$ to $2$ is
significant and the drop from depth $2$ to $3$ is not.

\begin{wrapfloat}{figure}[23]{r}[0pt]{0.68\textwidth}
\centering
\begin{subfigure}[t]{0.32\textwidth}
\centering
\includegraphics[width=\linewidth]{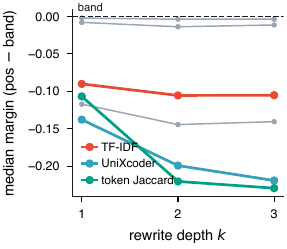}
\caption{Provenance evidence by depth}\label{fig:depth-margin}
\end{subfigure}\hfill
\begin{subfigure}[t]{0.32\textwidth}
\centering
\includegraphics[width=\linewidth]{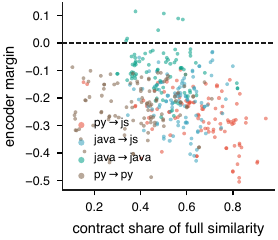}
\caption{Contract share versus encoder margin}\label{fig:depth-antagonism}
\end{subfigure}
\caption{(a) Median detector margin (positive score minus the per-source
band) over rewrite depth for three focal detectors, the remaining lexical
and encoder detectors in grey (the stat-profile channel lies below $-5$ at
all depths and is omitted from the panel), and the band as the dashed line.
(b) Contract share of the lexical cosine against the UniXcoder margin, one
point per instance on the benchmark's four rewriting axes
(\LinkSection{app:s2-contract}).}
\label{fig:rewrite-depth}
\end{wrapfloat}

\paragraph{Results.} Behavior is pinned: median conformance is $1.0$ at
every depth.
Median detector scores are already below the reference bands after the
first reconstruction
(\LinkFigure{fig:rewrite-depth}\hyperref[fig:rewrite-depth]{a}). Every
detector's median margin is negative at depth $1$ (tfidf $-0.090$,
token Jaccard $-0.107$, UniXcoder $-0.138$), and at most $6$ of $25$
sources still clear their band. The paired tests find an initial decline without a statistically resolved
additional decline: over chains complete at consecutive
depths, the drop from depth $1$ to $2$ is significant for tfidf
($p=0.020$), token Jaccard ($p=0.009$), and UniXcoder ($p=0.041$),
while the drop from depth $2$ to $3$ is not ($p \ge 0.075$), the
prespecified pattern used as a saturation diagnostic. Non-significance
of the later change does not establish equivalence or a true plateau. The deterministic channels echo the
same shape: string-literal margins fall from $-0.073$ to $-0.138$
between the first two rounds and then hold at $-0.139$, and identifier
margins drift without recovering ($-0.174$, $-0.255$, $-0.235$). Of the
$40$ chains, one stops at round $1$ when composition cannot recover
conformance; $23$ complete all three rounds, and $25$, $24$, and $23$
sources contribute an accepted artifact at depths $1$, $2$, and $3$.
Low margins after the first observed reconstruction support early loss of
detector-observable evidence. Later accepted rounds preserve the tested
behavior; failed and incomplete chains limit conclusions about the cost of
repeated rewriting.

\subsection{Same-language vs.\ cross-language provenance}
\label{app:language-pairing}
\label{app:t4-pairing}

\paragraph{Design.} Cross-language rewriting should remove more provenance
evidence than same-language rewriting, because language-specific lexical and
structural traces may change substantially under translation. We test this within sources:
$158$ release sources carry a verified artifact on both their same-language
and their cross-language axis ($132$ Python, $26$ Java), giving a paired
comparison that holds the source fixed. For each source and detector we take
the same-language margin minus the cross-language margin against the
source-specific band, with Wilcoxon signed-rank tests pre-registered before
computation.

\paragraph{Same-language rewriting retains a modest encoder-margin advantage.} Encoder margins are
higher on the same-language artifact for $74\%$ of paired sources
(\LinkTable{tab:language-pairing}). The median advantage is $+0.024$ for
Python pairs (Wilcoxon $p = 1.2\times10^{-6}$) and $+0.042$ for Java pairs
($p = 2.3\times10^{-5}$). TF--IDF shows no statistically resolved advantage
($p \geq 0.27$), so the encoder advantage is not mirrored by that lexical
measure. This does not identify the representation features responsible.
Median encoder margins remain below the band in both settings. The
corresponding median uncapped source-to-artifact size ratios are $0.010$ (py$\to$py) versus $0.012$ (py$\to$js) and $0.017$
(java$\to$java) versus $0.013$ (java$\to$js) on the paired sources.

\paragraph{Reading.} Cross-language rewriting has lower encoder margins
in the within-source comparison, consistent with the axis-level pattern in
\LinkSection{sec:main}. Holding sources fixed reduces composition effects,
but does not isolate translation from all other implementation changes.

\begin{table}[t]
\centering
\caption{Within-source paired comparison of same-language versus
cross-language artifacts on $158$ dual-axis sources. Margins are encoder
margins against the source-specific band (medians); $\Delta$ is the per-source
same-language minus cross-language margin.}
\label{tab:language-pairing}
\tablestyle\footnotesize
\begin{tabularx}{\linewidth}{Lrrrrr}
\toprule
\rowcolor{TableHeader}
\tablehead{Pair} & \tablehead{$n$} & \tablehead{Same} & \tablehead{Cross} & \tablehead{$\Delta$ median} & \tablehead{Wilcoxon $p$} \\
\midrule
Python ($132$ sources) & $132$ & $-0.151$ & $-0.175$ & $+0.024$ & $1.2\mathrm{e}{-6}$ \\
Java ($26$ sources) & $26$ & $-0.114$ & $-0.160$ & $+0.042$ & $2.3\mathrm{e}{-5}$ \\
\bottomrule
\end{tabularx}
\end{table}

\subsection{Repository scale and rewrite scope}
\label{app:scale-scope}
\label{app:t6-complexity}
\paragraph{Design.} Two pre-registered questions,
sealed before computation: does source
complexity explain where detection fails, and does restricting the source
to the slice its oracle actually covers recover encoder signal? Complexity
measures (LOC, file count, module count, import-graph edges, public-API
count) are computed statically from the pinned source clones. The
oracle-coverage slice of a source is the set of files touched by its
conformance harness plus their transitive imports; the definition was
fixed before any detector was run on the slice view.

\paragraph{Terciles.} Sources are split into complexity terciles within
each major axis (the source, not the instance, is the tercile unit), and
per-detector median margins and calibrated TPR are reported per tercile
with source-cluster bootstrap intervals. The result is
flat: UniXcoder's median margin stays between $-0.28$ and $-0.11$ in
\emph{every} tercile of \emph{every} measure on \emph{every} major
axis. The margin is slightly less negative for larger sources on the Python axes
(e.g.\ LOC terciles on py$\to$py: $-0.279$ low vs.\ $-0.230$ high) and
essentially unchanged on java$\to$java, where calibrated TPR never exceeds
$0.15$ in any tercile and is $0$ everywhere outside java$\to$java.
The gap persists across these complexity strata, so source complexity
alone does not explain the observed low calibrated detection.

\paragraph{Slice control.} Re-scoring UniXcoder with the source restricted
to its oracle-coverage slice (artifact unchanged, $88$ benchmark instances
with file-anchored slices) moves the positive score by a median of
$-0.033$, so restriction slightly \emph{lowers} similarity, and only $5/88$
instances cross the band that $0/88$ cross with the whole repository.
This control gives little support to source-file dilution as the sole
explanation. It does not exclude all effects of repository aggregation or
validate the slice as a complete semantic boundary.

\subsection{Generator-backbone ablation}
\label{app:generator-effects}

\begin{wrapfloat}{figure}{l}[0pt]{0.33\textwidth}
\vspace{-\baselineskip}
\centering
\includegraphics[width=\linewidth]{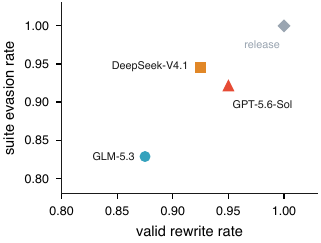}
\caption{\textbf{Generator-backbone ablation} under a fixed agentic rewriting
pipeline on 40 Python sources. We compare the default GPT-5.6-Sol backbone
with GLM-5.3 and DeepSeek-V4.1-Flash, varying only the generator backbone.
The $x$-axis reports the fraction of sources reaching full behavioral
conformance; the $y$-axis reports, among valid rewrites, the fraction below
the source-specific reference band on all seven static channels.
Released rewrites of the same sources (grey diamond) provide the reference
configuration; their validity is ensured by release selection.}
\label{fig:agentic-frontier}
\end{wrapfloat}

\paragraph{Generator backbone.}
GPT-5.6-Sol is the default generator backbone. We additionally evaluate
GLM-5.3 and DeepSeek-V4.1-Flash while holding the agentic rewriting
pipeline (\LinkSection{app:rewrite-protocol}), the 40 Python sources,
and the evaluation protocol fixed. GLM-5.3, DeepSeek-V4.1-Flash, and
GPT-5.6-Sol produce valid rewrite counts of 35/40, 37/40, and 38/40. Among
these valid outputs, 29/35, 35/37,
and 35/38 fall below the source band on all seven static channels, against 40/40
for the release rung over the same sources
(\LinkFigure{fig:agentic-frontier}), with wide margins (median UniXcoder
gap $-0.81$ versus $-0.27$ for release) and no embedding detector catching
any valid artifact. Frequent evasion of the seven static channels is
therefore not specific to a single tested generator backbone. This comparison does not
establish generator independence or a causal ordering by capability.

\subsection{Multi-source provenance}
\label{app:multisource}

The release benchmark covers single-source artifacts. As a first probe of
mixed-origin code, we construct composite artifacts by merging two
verified same-language rewrites of the full-rewrite rung, each parent
under its own top-level module directory. Pairing is deterministic over
hash-sorted sources, with mutual same-domain neighbours excluded, giving
$17$ Python and $14$ Java pairs. Each composite is scored against both
parents with the frozen lexical detectors on the canonical corpus, after
re-verifying all $189$ involved frozen source bands exactly.
Same-language non-parent sources serve as specificity controls.

Mixing dilutes per-parent signal. Under TF-IDF, the channel with the
strongest single-source margins, the mean margin drop from the
single-source reference to the composite is $0.119$, against a median
source band of $0.173$. Parents above their own band fall from $25/62$
for the single-source rewrites to $7/62$ for the composites, and both
parents clear their bands simultaneously in only $1/31$ pairs. The
effect is strongest in Python, where none of the parent scores remains above its band
($12/34$ parents above band before, $0/34$ after); in Java, $7/28$
parents remain above band, consistent with the stronger same-language
Java signal of \LinkTable{tab:axis-0}. The k-gram and statistical
profile channels already sit below band for single-source rewrites and
remain there. False positives remain: $6/155$ non-parent controls exceed their bands.

These file-level composites are harder to associate with both parents
under the evaluated whole-artifact scores. The result motivates testing
mixture-aware alternatives such as segment-level matching; it does not
establish that provenance signals can never compose. We leave spec-merged blends and mixture-aware detectors
to future work.

\section{Case Studies}
\label{app:case-studies}

This section makes the aggregate results legible through concrete
instances: what a rewrite that evades the seven static channels looks like, which
repositories resist evasion, and where the construction itself fails.
Every number comes from the frozen per-instance records released with the
benchmark and from the study materials cited inline.

\subsection{The complete agentic rewriting backend}
\label{app:case-backend}

The backend that re-executes the full pipeline below instantiates
\LinkAlgorithm{alg:repo-transformation}: every stage is a
separate headless coding-agent session with no shared context, driven
by a resumable scheduler that runs many pipelines in parallel.

\paragraph{Stages.} A \emph{decompose} session reads the source
repository and writes a unit inventory: three to eight core modules
ordered bottom-up, each specified by a responsibility, a precise
statement of observable behavior, the interface it must expose, and
the literal data strings its behavior fixes. File names, directory
layout, and the original project's identifiers are forbidden in the
inventory. Each unit is then rewritten by a \emph{fresh session in its
own directory} that receives only that unit's specification and the
interfaces of the units it depends on, never the source and never a
sibling implementation. A \emph{compose} session assembles the units
bottom-up, resolves interface mismatches by adjusting callers, adds
the conformance entry point, and iterates against the behavioral
oracle until conformance reaches
$1.0$: the candidate is copied to the evaluation host and run
against the source's captured golden cases at every iteration. The composed repository then becomes the input to the next
cycle: the decomposition is re-derived from scratch on the rewritten
code rather than inherited, so every cycle performs a fresh reconstruction while retaining the same
provenance path.
A final \emph{critic} session receives the source-chosen tokens that
survive in the composed repository and removes them by renaming
identifiers and rewording comments and messages, gated on conformance
staying $1.0$. Values the specification marks as externally required
(protocol field names, file-format keys) are kept: behavior wins over
leak removal. The exact stage prompts are reproduced in
\LinkAppendix{app:prompt}.

\paragraph{Models.} The same driver runs the identical pipeline with the
default GPT-5.6-Sol backbone and, for the backbone ablation, GLM-5.3 and
DeepSeek-V4.1-Flash against the same oracle.
The jwcrypto trace in \LinkAppendix{app:case-success} is one full
execution: six units decomposed, rewritten in isolation, and composed
to conformance $1.0$; a second cycle over the composed repository; a
critic pass; conformance $1.0$ (strict) throughout.

\subsection{What a successful evasion looks like}
\label{app:case-success}

\paragraph{The aggregate pattern.} Of the \nPair{} instances in the
frozen release suite, $87\%$ fall below the band of
all seven channels in this diagnostic: four lexical/statistical channels
and three encoders. This does not include the AST surrogate or native clone tools. Evasion is not confined to obscure repositories: every
py$\to$py and py$\to$js instance evades all
seven detectors, as do $96\%$ of java$\to$js instances. Detection lives
almost entirely on the java$\to$java axis, which supplies $87\%$ of the
detected instances (\LinkAppendix{app:java-tier}).

\begin{table}[t]
\centering
\caption{Three evading release artifacts (B5x critic rung): positive
score (band) per detector. All seven margins are negative on their native detector scales.}
\label{tab:case-evasion}
\tablestyle\footnotesize
\begin{tabularx}{\linewidth}{lLLL}
\toprule
\rowcolor{TableHeader}
\tablehead{Detector} & \tablehead{jwcrypto (py$\to$py)} & \tablehead{dask (py$\to$js)} & \tablehead{flink (java$\to$js)} \\
\midrule
tfidf & $0.188\,(0.291)$ & $0.029\,(0.203)$ & $0.017\,(0.103)$ \\
$k$-gram & $0.070\,(0.203)$ & $0.000\,(0.297)$ & $0.023\,(0.141)$ \\
token Jaccard & $0.086\,(0.320)$ & $0.002\,(0.390)$ & $0.000\,(0.022)$ \\
profile & $-9.57\,(-5.28)$ & $-26.6\,(-3.5)$ & $-27.2\,(-4.7)$ \\
UniXcoder & $0.814\,(0.948)$ & $0.617\,(0.857)$ & $0.506\,(0.861)$ \\
CodeBERT & $0.9956\,(0.9958)$ & $0.989\,(0.996)$ & $0.987\,(0.998)$ \\
GraphCodeBERT & $0.981\,(0.983)$ & $0.942\,(0.979)$ & $0.955\,(0.988)$ \\
\bottomrule
\end{tabularx}
\end{table}

\paragraph{Three cases.} \LinkTable{tab:case-evasion} makes the difference
between similarity and evidence concrete. Each source--rewrite score is
below its detector-specific band, even when absolute similarity is high.
For jwcrypto, CodeBERT scores the rewrite at $0.99560$ against a band of
$0.99580$, a margin of only $2\times10^{-4}$. The example illustrates a
threshold-dependent failure to distinguish derivation from the reference
code, not the disappearance of similarity. Native score scales differ, so
the numerical margins should not be ranked across detector families.

\paragraph{Which repositories resist evasion.} The repositories
with at least one detected instance are overwhelmingly small
java$\to$java utility libraries with ecosystem-standard vocabulary
(the Apache Commons family, \texttt{joda-time}, \texttt{snakeyaml},
\texttt{picocli}, \texttt{google-java-format}). These examples retain domain vocabulary in public interfaces, including
names such as \texttt{Duration} or \texttt{OptionParser}. Such vocabulary is
a plausible source of surviving similarity, but the cases do not identify
which tokens determine encoder scores. GraphCodeBERT
contributes to $82\%$ of the detections and CodeBERT to $58\%$;
\texttt{univocity-parsers} is the most
detected artifact in the benchmark (six of seven detectors).

\paragraph{Required behavior can retain recognizable evidence.}
On jwcrypto, whose release artifact sits below band on all seven static
channels, re-executing the full backend of \LinkAppendix{app:case-backend}
on the same source shrinks the surviving source vocabulary while behavior
stays intact: two complete cycles plus a critic pass cut surviving
source-chosen tokens from $143$ to $45$ while conformance stays
$1.0$. The surviving
vocabulary is the JOSE term set (JWK/JWS/JWE field names fixed by
RFC\,7515/7516) that the conformance cases compare against: the tested outputs constrain those strings. Required
behavior can thus retain recognizable evidence despite repeated reconstruction.

\subsection{When lexical evidence fails}
\label{app:case-clone-failure}

The lexical/statistical channels sometimes complement the encoders, but
their evidence is limited in these case-study records. Only $15\%$ of the detected instances are caught by a
classical detector that every encoder misses, and all eight are
java$\to$java (\texttt{java-jwt} and \texttt{joda-time} by tfidf,
\texttt{typetools} by tfidf and token Jaccard, one
\texttt{commons-digester} artifact by $k$-gram alone). The reverse
asymmetry is larger: \texttt{commons-text} (java$\to$java) is caught by
CodeBERT, GraphCodeBERT, tfidf and UniXcoder while its $k$-gram
positive is $0.031$ against a band of $0.086$ and its token Jaccard
$0.010$ against $0.057$. Once a
rewrite renames identifiers and reorders tokens, $n$-gram overlap falls
below the floor that unrelated repositories sharing common library
boilerplate already establish. These cases concern lexical/statistical similarities, not the provisional
SourcererCC, MSCCD, StoneDetector, TACC, or Oreo evaluations. They do not
establish dominance over native clone detectors.

\subsection{When an LLM / coding agent fails}
\label{app:case-llm-failure}

\paragraph{The source-visible camouflage recipe often fails behavioral checks.} In the independent
reimplementation ladder, the arm whose agent \emph{reads the full
source} and is asked to camouflage it in place produces artifacts that
often fail the behavioral gate: $14/29$ do not execute at all, and the best
conformance among those that run is $0.43$ (\texttt{django},
py$\to$js); \texttt{commonmark-java} runs at $0.0$. The arm that sees
only the behavioral contract reaches conformance $1.0$ on $24/29$
sources. These outcomes compare generation recipes with different inputs and
constraints; they do not show that access to source code itself causes
behavioral failure.

\paragraph{The tested mechanical interventions give limited valid evasion.} Five
mechanical interventions applied to $20$ release artifacts: identifier
splitting, merging, and string-literal renaming each destroy behavior
in $20/20$ artifacts; comment laundering preserves behavior in $16/20$
but moves the tfidf and $k$-gram margins by at most $0.009$; comment
stripping preserves behavior in $20/20$ with margin change exactly
$0$. The only margin that moves substantially ($+0.69$) is the profile
channel, the one channel the construction already controls. The tested primitives either break behavior or leave these lexical margins
nearly unchanged. This does not rule out other behavior-preserving
obfuscation methods.

\paragraph{Single-shot evasion varies by language axis.} The
strongest single-shot generator (\texttt{gpt-5.6-sol}, reimplementation
arm) produces artifacts that evade every detector on $15/21$ sources,
but on java$\to$java only $1/6$ evade; four of the five caught
java$\to$java artifacts are caught by CodeBERT and GraphCodeBERT.
The Java cases retain more detectable signal in this study, consistent
with the examples in \LinkAppendix{app:case-success}. This does not establish
a general relationship between generator capability and evasion.

\subsection{Multi-source provenance}
\label{app:case-multisource}

Every artifact in the frozen benchmark descends from exactly one
source repository; multi-ancestor artifacts (blends of two sources, or
a rewrite seeded from a competitor's implementation) are outside the
current construction. The pair task's negative sets already probe the
boundary: each source is scored against its same-domain competitors
(e.g.\ \texttt{jwcrypto} against \texttt{joserfc}, \texttt{pyjwt},
\texttt{pyjwkest}, \texttt{jwskate}), so a detector must separate
derivation from same-domain independent implementation. A first
small-scale probe that merges two verified rewrites at the file level
finds that per-parent provenance signal is diluted below the source
bands, so both parents of a composite are rarely recovered together
(\LinkAppendix{app:multisource}). Constructing blended artifacts by
decomposing two sources and merging their specifications remains a future extension of the pipeline, distinct from this file-level
composition probe.

\subsection{Failure cases of benchmark construction}
\label{app:case-failures}

\paragraph{Purpose-only specifications under-determine behavior.} The
ladder arm whose agent receives only a one-paragraph purpose statement
produces no artifact that reaches the $0.99$ conformance gate ($0/29$);
five do not execute, and the conformances that run span $0.0$--$0.5$.
A purpose statement fixes what a system is \emph{for} but not what it
\emph{does}; a purpose statement alone is insufficient under this study's strict gate.

\paragraph{Contracts can under-determine behavior too.} Even the
interface-only arm fails on sources whose observable behavior is not
captured by the extracted contract: \texttt{commons-digester} reaches
only $0.375$, \texttt{commons-dbutils} $0.82$, \texttt{jsoup} $0.85$.
The oracle, not the contract, is the ground truth; instances whose
goldens exercise behavior outside the contract are exactly where
independent reimplementation falls short, and they are excluded by the
study's strict verification gate. This is distinct from the release
criterion, which permits partial conformance.

\paragraph{Purpose descriptions leak generic vocabulary.} The purpose
statements carry $0$--$45$ source-chosen tokens per source, but
inspection shows they are generic domain words (\texttt{admin},
\texttt{manage}, \texttt{preview}) rather than identifying names; we
disclose the counts rather than claim zero leakage.

\paragraph{Interventions can backfire.} The comment-laundering
intervention of \LinkAppendix{app:case-llm-failure} moves the profile
margin by $+0.69$. Under the score-minus-band convention, a positive
change moves toward detection, not away from it. The profile channel is
sensitive to the altered implementation statistics; this intervention does
not establish that those statistics uniquely identify provenance.

\section{Robustness and Limitations}
\label{app:robustness}

\subsection{Metrics, calibration, and statistical units}
\label{app:metrics-calibration}
\label{app:metrics}
\paragraph{Independent units.}
Pairwise rows are repository-pair observations clustered by generating source;
attribution rows are query repositories with fixed candidate sets. Rewrites of
one source are not treated as independent source replicates. Pooled descriptive
metrics use all available instances, while uncertainty checks resample sources
and retain all their associated instances. All intervals are conditional on
the frozen sources, generated artifacts, negative neighborhoods, and task manifests; they do not represent uncertainty over a new generation run.

\paragraph{Pairwise and attribution metrics.}
AUROC and best F1 summarize threshold-free ordering and the best retrospective
threshold, respectively; best F1 is not a deployable operating point. For
detector $d$ and source $s$, the independent-code band is
$\beta_d(s)=\max_{n\in\mathcal{N}(s)}S_d(s,n)$. Calibrated TPR and FPR apply
the strict test $S_d(q,c)>\beta_d(s)$ to positive and same-domain negative
pairs. Rank@1 gives fractional credit when the true source is tied at the top.
Calibrated attribution additionally requires that top source to clear its band;
band passage alone is not attribution. Missing parses, unsupported languages,
timeouts, and empty eligible views remain unavailable outcomes rather than
numerical zeroes.

\begin{table}[t]
\centering
\caption{What each detector family measures, and the failure exposed by the
benchmark. This table interprets, rather than repeats, the numerical suite in
\LinkTable{tab:suite}.}
\label{tab:family-failures}
\tablestyle
\begin{tabularx}{\linewidth}{lLL}
\toprule
\rowcolor{TableHeader}
\tablehead{Family} & \tablehead{Evidence used} & \tablehead{Observed failure mode} \\
\midrule
Lexical/statistical & Tokens, character substrings, repository shape & Reconstruction reduces incidental overlap; shared contract vocabulary and style can also raise non-source scores \\
Token clone & Indexed file/block overlap & Near-random discrimination on readable inputs, with incomplete repository and language coverage \\
Structural/learned clone & Parsed functions, AST regions, metrics, pair classifiers & Provisional native results require parsing, mapping, and band validation; the AST surrogate has similar pooled TPR and FPR \\
Repository embedding & Pooled pretrained window representations & Relative ordering survives more often than calibrated attribution; evidence varies by language axis \\
\bottomrule
\end{tabularx}
\end{table}

\LinkTable{tab:family-failures} connects those metric definitions to the
distinct applicability and decision failures observed for each detector family.

\paragraph{Uncertainty and multiplicity.}
The primary benchmark reports descriptive task metrics rather than treating
every detector--axis cell as an independent hypothesis test. For calibrated
success rates, 4,000 source-cluster bootstrap resamples provide percentile
95\% intervals; for example, Pair TPR is $0.023$ for UniXcoder
($[0.004,0.048]$) and $0.074$ for CodeBERT ($[0.035,0.120]$). Targeted paired
studies define the source as the unit, report the paired effect and source
bootstrap interval, and use two-sided Wilcoxon tests only for their sealed
primary comparison. The full per-axis record, control tier, strict-positive
sensitivity, and native-tool applicability audit follow below.

\subsubsection{Sensitivity to calibration neighbors and missing comparisons}
\label{app:neighbor-sensitivity}
\paragraph{Single-neighbor deletion.}
We use the AST surrogate's unrounded pair scores to
perturb only the calibration neighborhood. For a source with at least two
neighbors, define
\[
\beta_d^{-}(R_s)=\min_{N\in\mathcal{N}(R_s)}
\max_{M\in\mathcal{N}(R_s)\setminus\{N\}}d(R_s,M).
\]
This is the smallest band obtainable by deleting one neighbor, equivalently
the second-largest source--neighbor score, counting repeated maxima. It
measures the strongest single-deletion effect, not an average over deletions.
Artifacts, task membership, negative pairs and attribution candidates remain
fixed; the deleted calibration neighbor remains in the evaluation set.
The deletion is determined by source--neighbor scores, not query outcomes.
We exclude 17 Pair and four Attr queries with singleton neighborhoods from
both columns, since deletion leaves their band undefined.

\begin{table}[t]
\centering
\caption{AST-surrogate calibration sensitivity on the same eligible queries in both
columns. Pair contains 417 queries from 111 sources and 2,037 negative pairs;
Attr contains 96 queries from 80 sources. Attr uses conjunctive top-tie credit.
No detector was rerun and no benchmark instance was changed.}
\label{tab:neighbor-sensitivity}\tablestyle
\begin{tabularx}{\linewidth}{Lrr}
\toprule
\tablehead{Outcome} & \tablehead{Original band} & \tablehead{Single-deletion minimum} \\
\midrule
Pair true positives & 46/417 & 71/417 \\
Pair false positives & 255/2,037 & 354/2,037 \\
Attr success credit & 1/96 & 2/96 \\
\bottomrule
\end{tabularx}
\end{table}

The band decreases for 404 Pair and 91 Attr queries. More positives cross
the lower band, but so do more negatives (\LinkTable{tab:neighbor-sensitivity}).
This sensitivity does not validate the neighbors' independent origin or
validate published TreeCen or provide a candidate-available prediction rule: it retains the generating
source's grouped band. The other six channels' rounded per-instance
summaries are not used to reconstruct this perturbation. The frozen encoder
records behind \LinkTable{tab:suite} store each source's band but not its
individual source--neighbor scores, so this sensitivity has not been
computed for the repository encoders. A source-isolated
calibration and source-absent evaluation remain needed.

\paragraph{StoneDetector records without flagged missing comparisons.}
Extending the timeout-only exclusion in \LinkAppendix{app:stone-audit}, we
exclude a query if any required positive, negative/candidate or
source--neighbor comparison is marked error or unsupported. This leaves
60 Java-to-Java Pair queries from 22 sources, with TPR $10/60$ and FPR
$5/248$, and 14 Attr queries from 14 sources, with conjunctive success
$0/14$. These are conditional summaries of records without those flags,
not performance estimates for omitted queries. Absence of a flag does not
certify parsing or a valid no-match result; self-copy, partial-copy and
local-coverage checks remain unperformed. The provisional record is the one
of \LinkAppendix{app:stone-audit}.
The released audit materials record the input
hashes and a read-only standard-library reproduction of both analyses.

\subsection{Negative/candidate-pool sensitivity}
\label{app:pool-sensitivity}

\begin{figure*}[t]
\centering
\begin{minipage}[t]{0.32\textwidth}
  \centering
  \includegraphics[width=\textwidth]{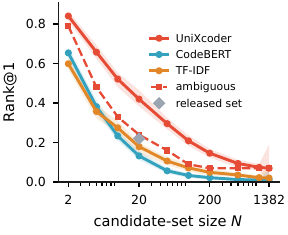}\\[2pt]
  \textbf{(a)} ranking decay
\end{minipage}\hfill
\begin{minipage}[t]{0.32\textwidth}
  \centering
  \includegraphics[width=\textwidth]{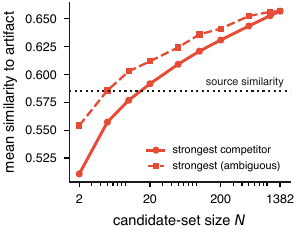}\\[2pt]
  \textbf{(b)} extreme-value effect
\end{minipage}\hfill
\begin{minipage}[t]{0.32\textwidth}
  \centering
  \includegraphics[width=\textwidth]{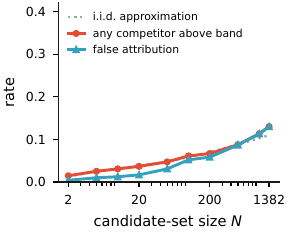}\\[2pt]
  \textbf{(c)} calibrated failure
\end{minipage}
\caption{Candidate-pool sensitivity with candidate-set size $N$ on a log scale. (a) Tie-aware Rank@1 decays smoothly as $N$ grows, both for
random competitor draws (solid) and for domain-ambiguous draws that keep the
$N-1$ pool repositories most similar to the source (dashed, strongest
detector). The grey diamond marks the released 20-candidate task.
(b) The extreme-value mechanism behind the decay: the mean similarity of the
strongest independent competitor grows with $N$ while residual
artifact-to-source similarity (dotted) is constant in $N$; the two cross
already near $N{\approx}20$.
(c) The empirical rate of drawing at least one above-band competitor is
shown alongside the independent-draw approximation $1-(1-p)^{N-1}$, where
$p$ is the per-query fraction of pool competitors above the frozen source
band. This expression is not a union bound. The false-attribution curve
uses the study's fixed-band rule and does not establish deployment error
control.}
\label{fig:attr-scaling}
\end{figure*}

\paragraph{Larger candidate pools create stronger competing matches.}
The primary task fixes 20 candidates per query. In the pool-size sensitivity,
UniXcoder's Rank@1 under random competitor draws falls from 0.84 at $N=2$
through 0.42 at $N=20$ to 0.07 over the full 1,382-repository pool
(\LinkFigure{fig:attr-scaling}a). This does not replace the primary Rank@1
of 0.220: the random draws and the fixed benchmark candidate sets differ.
Source similarity is fixed while the strongest competitor score increases
with pool size (\LinkFigure{fig:attr-scaling}b), explaining how retrieval can
worsen without a change in the rewrite itself.

The chance of drawing an above-band competitor also grows
(\LinkFigure{fig:attr-scaling}c). The dotted curve is an independent-draw
approximation; a general union bound would instead be
$\min\{1,(N-1)p\}$. The fixed source band does not imply a constant
operational abstention rate, because that rate also depends on candidate
scores and the prediction rule. These curves stress-test candidate
competition, rather than validate open-set attribution or query-level error
control. The fixed 20-candidate task is harder than random draws at the same
size, consistent with the role of its same-domain distractors.

\subsection{Additional validity audits}
\label{app:validity-audits}

\paragraph{Does the acceptance gate test behavior or formatting?} The
behavioral oracle compares program outputs, so acceptance could in principle
hinge on superficial formatting rather than computation. We audit this on the
generator-frontier study of \LinkSection{app:generator-effects} by classifying
every conformance case ($1{,}011$ cases over $40$ sources) into core cases,
which check computational behavior that the specification requires, and
surface cases, which check formatting, persona, or greeting-style output. A
strong model performs the classification and every doubtful case defaults to
core. The suite is almost purely behavioral: $1{,}005$ cases are core and only
$6$ are surface, all in two sources whose contracts involve conversational or
human-facing output (\texttt{ChatGLM-6B} and \texttt{TagStudio}).

\paragraph{Relaxed-gate sensitivity.} We then re-gate the study rewrites with
a two-tier oracle that keeps exact comparison on core cases but checks only
exit status, error class, and output type structure on surface cases. All
$37$ valid rewrites in the evaluated generator arm still pass at conformance $1.0$. This holds even after an
agent pass that rewrites the surface outputs themselves: persona and branding
strings disappear, the count of surviving source-chosen tokens drops from
$740$ to $539$, and core behavior is untouched. The longest core-case outputs
are computational renderings such as markup escaping, table layout, and
changelogs, not stylistic prose.

\paragraph{Reading.} Core computational cases dominate this generator study.
Relaxing the six surface cases preserves its validity outcomes. This supports
the behavioral role of the oracle in this study, without establishing the
same case composition for every benchmark source.

\subsection{Limitations}
\label{app:limitations}
\label{app:gen-limitations}
The behavioral oracle checks a finite set of deterministic behaviors
visible to the generator, so passing it establishes oracle-covered
behavior, not whole-repository equivalence; the benchmark contract allows partial
positive-case conformance. The bounded unseen-input audit of
\LinkSection{app:functional-preservation} uses benchmark instances. Its inputs were
authored by this project's own agent rather than an independent party, and
two original-side environment repairs were applied before execution. The
audit therefore covers the stated contract rather than whole repositories.
The adapter token limit excludes degenerate artifacts but does not prove that
a thin adapter performs only interface mapping. Oracle and environment
requirements also determine which repository behaviors can be evaluated.

\end{document}